# Transverse and longitudinal angular momenta of light

Konstantin Y. Bliokh[1,2] and Franco Nori[1,3]

[1]*Center for Emergent Matter Science, RIKEN, Wako-shi, Saitama 351-0198, Japan*
[2]*Nonlinear Physics Centre, RSPhysE, The Australian National University, Canberra, ACT 0200, Australia*
[3]*Physics Department, University of Michigan, Ann Arbor, Michigan 48109-1040, USA*

We review basic physics and novel types of optical angular momentum. We start with a theoretical overview of momentum and angular momentum properties of generic optical fields, and discuss methods for their experimental measurements. In particular, we describe the well-known *longitudinal* (i.e., aligned with the mean momentum) spin and orbital angular momenta in polarized vortex beams. Then, we focus on the *transverse* (i.e., orthogonal to the mean momentum) spin and orbital angular momenta, which were recently actively discussed in theory and observed in experiments. First, the recently-discovered *transverse spin* angular momenta appear in various *structured* fields: evanescent waves, interference fields, and focused beams. We show that there are several kinds of transverse spin angular momentum, which differ strongly in their origins and physical properties. We describe extraordinary features of the transverse optical spins and overview recent experiments. In particular, the helicity-independent transverse spin inherent in edge *evanescent waves* offers robust spin-direction coupling at optical interfaces (the quantum spin Hall effect of light). Second, we overview the *transverse orbital* angular momenta of light, which can be both *extrinsic* and *intrinsic*. These two types of the transverse orbital angular momentum are produced by *spatial shifts* of the optical beams (e.g., in the spin Hall effect of light) and their *Lorentz boosts*, respectively. Our review is underpinned by a unified theory of the angular momentum of light based on the *canonical* momentum and spin densities, which avoids complications associated with the separation of spin and orbital angular momenta in the Poynting picture. It allows us to construct comprehensive classification of all known optical angular momenta based on their key parameters and main physical properties.

## 1. Introduction

   Angular momentum (AM) was recognized as one of the important characteristics of light after the pioneering work by Poynting [1] and the first experimental evidence by Beth [2]. These works considered the *spin* AM produced by the *circular polarization* of a light beam. In 1992, a paper by Allen *et al*. [3] started a new era of AM studies in optics. This work described the *orbital* AM in so-called *vortex* beams, which was soon detected experimentally [4]. Since then, study of optical angular momentum has grown into a large research field with numerous applications in optical manipulations, quantum information, photonics, plasmonics, and astrophysics (see books [5–9] and reviews [10–14]).

   Importantly, the *spin* and *orbital* AM of light are separately observable properties in optics [5–20]. The separation of AM into spin and orbital parts is straightforward in paraxial monochromatic beams [3,5–14,21]. At the same time, fundamental difficulties in quantum electrodynamics and field theory [22–26] result in a number of subtleties for the spin and orbital AM description in generic non-paraxial or non-monochromatic fields [27–38]. Still, the spin-orbital AM decomposition is possible and physically meaningful in optics, and we will rely on it in this review.

The spin AM is associated with the *polarization* of light, so that right-hand and left-hand *circular polarizations* of a paraxial beam correspond to the positive and negative *helicities* $\sigma = \pm 1$ of photons. From the paraxial-beam optics [1–18,21] and the quantum mechanics of photons [22], we know that the spin AM of light is aligned with the direction of propagation of light. If the mean momentum of the beam (in units of $\hbar$ per photon) can be associated with its mean wave vector $\langle \mathbf{k} \rangle$, then such beam carries the corresponding spin AM $\langle \mathbf{S} \rangle = \sigma \langle \mathbf{k} \rangle / k$. Here $\sigma$ is the *helicity* parameter, i.e., the degree of circular polarization.

In turn, the orbital AM of a uniformly-polarized paraxial beam is independent of the polarization, and is produced by the *phase* and momentum circulation in a scalar wave function. For instance, an *optical vortex*, i.e., a phase singularity with a helical phase around it [39–43], generates a circulation of the local momentum density (phase gradient) [14,30] and, thereby, produces the orbital angular momentum. The orbital AM states of monochromatic light are paraxial vortex beams, which carry the orbital AM along the beam axis: $\langle \mathbf{L} \rangle = \ell \langle \mathbf{k} \rangle / k$ [3–18,21]. Here $\ell$ is the quantum number (topological charge) of the optical vortex.

Thus, in the majority of situations considered so far [1–21], the angular momentum of light is aligned with its mean momentum $\langle \mathbf{k} \rangle$. In other words, it is *longitudinal*. Recently, there has been a rapidly growing interest in optical fields with a *transverse* AM [44–59]. Such angular momentum is orthogonal to the propagation direction (mean momentum) of light. Importantly, different transverse angular momenta, of either spin or orbital nature, have drastically different physical properties and origins.

For example, the simplest case of the transverse AM is the orbital AM of a classical point particle with the coordinates $\mathbf{r}$ and momentum $\mathbf{p}$: $\mathbf{L} = \mathbf{r} \times \mathbf{p}$ [60]. This AM is *extrinsic*, i.e., it depends on the choice of the coordinate origin. Similar extrinsic orbital AM is also carried by a light beam passing at some distance from the coordinate origin: $\langle \mathbf{L}^{\text{ext}} \rangle = \langle \mathbf{r} \rangle \times \langle \mathbf{k} \rangle$, where $\langle \mathbf{r} \rangle$ is the mean position, i.e., *shift* of the beam. Despite its "trivial" character, in 1987 Player and Fedoseyev showed [61,62] that this extrinsic orbital AM plays a key role in the transverse spin-dependent shifts of light beams reflected or refracted at planar interfaces. The latter effect is now known as the *spin-Hall effect* of light or Imbert–Fedorov shift [61–73], and the extrinsic transverse orbital AM was intensively considered in this context (see [73] for a review).

In contrast to the previous example, there is a more intriguing *transverse spin AM*, which was first described in 2012 by Bliokh and Nori [46]. This transverse spin appears locally in *structured* optical fields, such as evanescent waves, focused beams, and two-wave interference [46,48,50–59]. The transverse spin has extraordinary properties, which are in sharp contrast to what has been known so far about the spin of photons. Namely, it is *independent of the helicity* of the wave and can appear even for *linearly* polarized waves. (Actually, here we show that there is a family of transverse-spin AM, which have different physical properties and depend on different field parameters.) Moreover, this polarization-independent transverse spin strongly depends on the *direction of propagation* of the wave. Owing to such extraordinary features, the transverse spin AM in evanescent waves has already found important applications in spin-dependent unidirectional optical interfaces [52,53,56,57,74–83]. Remarkably, the robust coupling of the transverse spin with the propagation direction in evanescent waves is a fundamental property of Maxwell equations, which can be associated with the *quantum spin Hall effect* of light [56].

Here we overview recent theoretical and experimental investigations of different types of AM of light. By exploiting a unified theory of optical angular momentum based on the *canonical* (rather than Poynting) momentum and spin densities [34,37,38,50,54,84], we aim to provide a coherent picture of all the basic kinds of AM, and describe their main physical features. We particularly focus on the *transverse* AM, which so far have not been considered in reviews on the optical AM [5–14]. We describe basic systems, where the transverse AM may appear and play an important role, and discuss the main experimental works related to the transverse AM (even where its presence was not recognized).

The paper is organized as follows. In Section 2 we provide an overview of the general AM properties in mechanics and wave physics. We describe canonical momentum, angular-momentum, spin, and helicity densities in generic optical fields, consider paraxial optical beams carrying longitudinal spin and orbital AM, and discuss measurements of the spin and orbital AM in optical fields. Section 3 considers various types of the transverse spin AM in structured (inhomogeneous) fields. These include evanescent waves, two-wave interference, and focused beams. In addition to a detailed theoretical analysis, we review a number of experimental works revealing the transverse spin. The relation of the transverse spin in evanescent waves to the spin-direction coupling (quantum spin Hall effect of light) is discussed. Section 4 examines two types of transverse orbital AM: the extrinsic and intrinsic ones. We show that the former (extrinsic) case is related to transverse shifts of paraxial beams, and it plays an important role in the spin Hall effect of light. The latter (intrinsic) case is realized in polychromatic spatio-temporal beams, which can be obtained, e.g., via a transverse Lorentz boost to a moving reference frame. Section 5 summarizes the results and provides a classification of different kinds of AM (Table I).

*1.1. Conventions and notations*

Throughout this paper, different types of the AM are emphasized by frames around their equations, and their key properties (spin/orbital, intrinsic/extrinsic, longitudinal/transverse, etc.) are highlighted in special *framed equations with Roman numbers*: (I), (II), etc.

It should be emphasized that "transverse" with respect to the mean momentum can be applied to *two* mutually orthogonal directions. In many problems, the structured waves are formed by at least two wave-vector or momentum quantities forming the "plane of propagation" (as, e.g., in an evanescent wave or two-wave interference). In these cases, we distinguish "*transverse (out-of-plane)*" and "*transverse (in-plane)*" quantities, which typically strongly differ in their properties and origin.

We pay special attention to the *time-reversal* ($\mathcal{T}$) and *spatial-inversion* ($\mathcal{P}$) *symmetries* of the parameters determining different types of angular momentum. Any AM must change its sign under the time-reversal ($\mathcal{T}$) transformation, but preserve it under the spatial-inversion ($\mathcal{P}$) transformation. This enables a clear identification and separation of different kinds of AM, based on the $\mathcal{P}$- and $\mathcal{T}$-symmetries of the key parameters in their equations.

Overall, the framed equations throughout the paper and the *final Table I* should provide a clear guide to and summary of the main results.

We also remark on the notations and units used in this review. First, throughout the paper (apart from Section 4.2) we deal with *monochromatic* electromagnetic fields of fixed frequency $\omega$, characterized by complex electric and magnetic field amplitudes $\mathbf{E}(\mathbf{r})$ and $\mathbf{H}(\mathbf{r})$. The real electric and magnetic fields are given by $\mathcal{E}(\mathbf{r},t) = \text{Re}\left[\mathbf{E}(\mathbf{r})e^{-i\omega t}\right]$ and $\mathcal{H}(\mathbf{r},t) = \text{Re}\left[\mathbf{H}(\mathbf{r})e^{-i\omega t}\right]$.

Second, for the sake of simplicity, we mostly consider optical fields in the *vacuum* and use *Gaussian units* (to avoid multiple appearances of the vacuum permittivity and permeability constants). Furthermore, since we do not consider any truly quantum phenomena, we imply $\hbar = 1$ units when using the quantum operator formalism for the momentum and AM of light. In this manner, the energy and momentum of a wave are directly associated with the frequency $\omega$ and wave vector $\mathbf{k}$.

Finally, to make the main equations as simple and clear as possible, we often omit inessential factors and use the proportionality sign "$\propto$" instead of the exact equality sign "$=$".

## 2. Spin and orbital angular momenta: basic properties

### *2.1. Particles and waves*

We start with an introductory discussion of the AM properties in mechanics and wave physics. The AM reveals very different features for particles and waves. First, let us consider the limiting case of a classical *point particle*. Such particle carries *orbital AM* [60]

$$\boxed{\mathbf{L} = \mathbf{r} \times \mathbf{p}}, \tag{2.1}$$

where $\mathbf{r}$ and $\mathbf{p}$ are the position and momentum of the particle, respectively (Fig. 1a). This mechanical AM is *extrinsic*, i.e., dependent on the choice of the coordinate origin. Indeed, the translational transformation of the coordinates,

$$\mathbf{r} \to \mathbf{r} + \mathbf{r}_0, \tag{2.2}$$

changes the AM (2.1) as

$$\mathbf{L} \to \mathbf{L} + \mathbf{r}_0 \times \mathbf{p}. \tag{2.3}$$

By definition, a point particle cannot have any internal structure and, hence, cannot carry an intrinsic AM. Evidently, the AM (2.1) is *transverse*, i.e., orthogonal to the momentum. Since the momentum $\mathbf{p}$ is a $\mathcal{P}$-odd and $\mathcal{T}$-odd quantity, whereas the position $\mathbf{r}$ is $\mathcal{P}$-odd and $\mathcal{T}$-even, the AM (2.1) $\mathbf{L}$ is $\mathcal{P}$-even and $\mathcal{T}$-odd, as it should be. Thus, summarizing the properties of the point-particle AM:

> **Point particle AM:** Orbital, Extrinsic, Transverse. Key parameters: $\mathbf{p}, \mathbf{r}$. (I)

In contrast to localized particles, waves are extended entities. The limiting case of a wave entity is a *plane wave*. Such wave cannot carry an extrinsic or orbital AM of the type (2.1), because its position is undefined. A circularly-polarized electromagnetic plane wave, propagating along the $z$-axis can be written as

$$\mathbf{E} \propto \frac{\overline{\mathbf{x}} + i\sigma \overline{\mathbf{y}}}{\sqrt{2}} \exp(ikz), \quad \mathbf{H} = \overline{\mathbf{z}} \times \mathbf{E}, \tag{2.4}$$

where $\overline{\mathbf{x}}$, $\overline{\mathbf{y}}$, and $\overline{\mathbf{z}}$ denote the unit vectors of the corresponding Cartesian axes, the *helicity* parameter $\sigma = \pm 1$ corresponds to the right-hand and left-hand circular polarizations (Fig. 1b), and the wave number $k = \omega/c$. The electric (or magnetic) field (2.4) represents eigenmodes of the $z$-component of the spin-1 matrix operator $\hat{\mathbf{S}}$ (generators of the SO(3) vector rotations) [21,22,31]:

$$\hat{\mathbf{S}} = -i \left\{ \begin{pmatrix} 0 & 0 & 0 \\ 0 & 0 & 1 \\ 0 & -1 & 0 \end{pmatrix}, \begin{pmatrix} 0 & 0 & -1 \\ 0 & 0 & 0 \\ 1 & 0 & 0 \end{pmatrix}, \begin{pmatrix} 0 & 1 & 0 \\ -1 & 0 & 0 \\ 0 & 0 & 0 \end{pmatrix} \right\}, \quad \hat{S}_z \mathbf{E} = \sigma \mathbf{E}. \tag{2.5}$$

Therefore, the plane wave (2.4) carries the *spin AM density* $\mathbf{S}$ defined as the *local* expectation value of the operator $\hat{\mathbf{S}}$ with the wavefunction $\mathbf{E}$ or $\mathbf{H}$ [21,34] (see also Sections 2.2 and 2.3):

$$\boxed{\mathbf{S} \propto \sigma \frac{\mathbf{k}}{k}}. \tag{2.6}$$

Here we wrote the spin density in a form valid for an arbitrary propagation direction of the plane wave, $\mathbf{k}/k$.

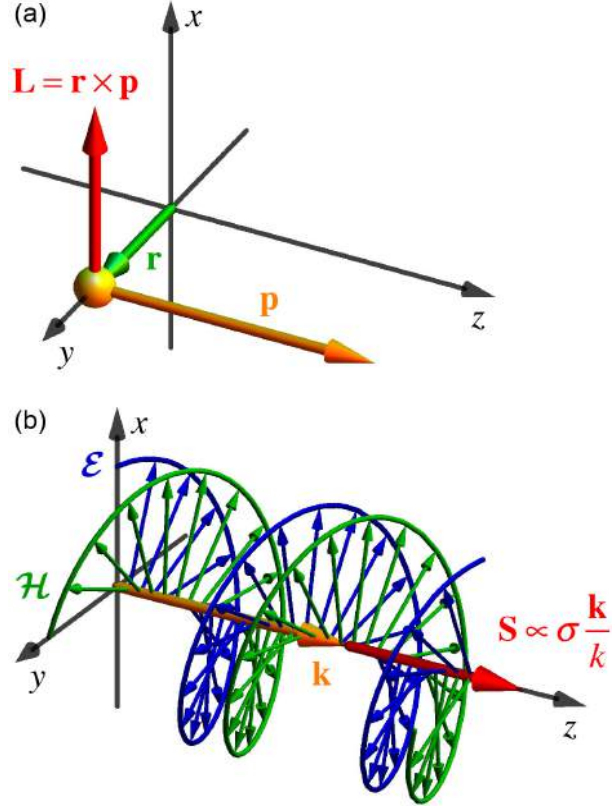

**Fig. 1.** Angular momentum of pure particles and waves. **(a)** A classical point particle carries orbital AM (2.1) and (I) $\mathbf{L}$, which is extrinsic and transverse (i.e., orthogonal to the particle momentum $\mathbf{p}$). **(b)** A single electromagnetic plane wave carries spin AM density (2.6) $\mathbf{S}$, which is intrinsic and longitudinal (i.e., aligned with the wave vector $\mathbf{k}$). This spin AM is produced by the rotation of the electric and magnetic fields in a circularly-polarized wave. Here the instantaneous distribution of the real electric and magnetic fields $\mathcal{E}(z,0)$ and $\mathcal{H}(z,0)$ is shown for a right-hand circularly-polarized wave with helicity $\sigma = 1$.

The spin AM density (2.6) is generated by rotating electric and magnetic fields in a circularly-polarized wave, Fig. 1b. The total (integral) AM is ill-defined for a plane wave, as it cannot be localized and all integrals diverge. Evidently, the spin density (2.6) is independent of the radius-vector $\mathbf{r}$, and thus it is an *intrinsic* quantity. Furthermore, this AM is *longitudinal*, i.e., directed along the momentum (wave vector) $\mathbf{k}$. The helicity $\sigma$ represents the chirality of light [85–89], and, hence, it is a $\mathcal{P}$-odd and $\mathcal{T}$-even quantity. Since the wave momentum $\mathbf{k}$ is $\mathcal{P}$-odd and $\mathcal{T}$-odd, the spin (2.6) has the proper $\mathcal{T}$-odd symmetry of the AM. Summarizing the properties of the plane-wave AM:

> **Plane wave AM:** Spin, Intrinsic, Longitudinal. Key parameters: $\sigma$, $\mathbf{k}$. (II)

Thus, we see that the AM of point particles and plane waves exhibit drastically different properties. These limiting cases encompass the variety of AM features appearing in *structured* optical fields.

Since optics deals with waves rather than particles, it is natural to look at the *quantum* counterpart of the mechanical orbital AM (2.1). It is described by the canonical quantum operator $\hat{\mathbf{L}} = \hat{\mathbf{r}} \times \hat{\mathbf{p}}$ [21,22,31], where $\hat{\mathbf{r}} = \mathbf{r}$ and $\hat{\mathbf{p}} = -i\nabla$ in the coordinate representation. Using the cylindrical coordinates $(\rho, \varphi, z)$, one can write the $z$-component of this operator and its eigenmodes as [3–14,21,31]:

$$\hat{L}_z = -i\frac{\partial}{\partial \varphi}, \quad \mathbf{E} \propto \exp(i\ell\varphi), \quad \hat{L}_z \mathbf{E} = \ell \mathbf{E}. \tag{2.7}$$

These eigenmodes are *vortices*, i.e., helical-phase waves, where $\ell = 0, \pm 1, \pm 2, \ldots$ is the vortex charge [39–43]. The vortex modes essentially represent *structured* wave fields (they are produced by the interference of multiple plane waves), and, hence, combine some properties of particles and waves. We consider the simplest examples of such modes, paraxial vortex beams, in Section 2.3. Remarkably, in sharp contrast to the mechanical orbital AM (2.1), the orbital AM of the vortex beams becomes *longitudinal* and even *intrinsic*.

## *2.2. Fundamental properties of generic optical fields*

We now describe the main momentum and angular-momentum properties of *generic* optical fields. We still consider monochromatic fields with frequency $\omega$, but they can have arbitrarily inhomogeneous spatial distributions $\mathbf{E}(\mathbf{r})$ and $\mathbf{H}(\mathbf{r})$, which fulfill free-space Maxwell equations.

Prior to describing angular momentum, we need to introduce the energy and momentum of light. The time-averaged energy density of a monochromatic optical field is [90,91]

$$W = \frac{g\omega}{2}\left(|\mathbf{E}|^2 + |\mathbf{H}|^2\right), \tag{2.8}$$

where $g = (8\pi\omega)^{-1}$ in Gaussian units.

It is widely accepted that the momentum density of a free electromagnetic field is described by the Poynting vector $\mathbf{\Pi} = gk\,\mathrm{Re}(\mathbf{E}^* \times \mathbf{H})$ [90,91]. However, it turns out that the more relevant and directly measurable quantity is the so-called *canonical (or orbital) momentum* density [14,30,34,50,54,89,92], which appears in canonical Noether conservation laws (in the Coulomb gauge) in electromagnetic field theory [23,34,37,38]. This momentum density can be written as

$$\mathbf{P} = \frac{g}{2}\,\mathrm{Im}\left[\mathbf{E}^* \cdot (\nabla)\mathbf{E} + \mathbf{H}^* \cdot (\nabla)\mathbf{H}\right], \tag{2.9}$$

where we use the notation $\mathbf{X} \cdot (\mathbf{Y})\mathbf{Z} \equiv \sum_i X_i Y Z_i$. Unlike the Poynting vector, the canonical momentum density (2.9) has an intuitively-clear physical interpretation: it is proportional to the local gradient of the phase of the field, i.e., to the *local wave vector* [30] (assuming contributions from both the electric and magnetic fields). Therefore, the canonical momentum density is independent of the polarization in uniformly-polarized fields and can be equally defined for a scalar wave field $\psi(\mathbf{r})$. The Poynting vector coincides with the canonical momentum in all cases where one can neglect elliptical polarization and spin AM phenomena. In all other situations these differ by the so-called "spin momentum" contribution introduced by Belinfante in field theory and non-measurable in basic light-matter interactions [14,23,25,30,34,50,54].

The circulation of the canonical momentum immediately yields the *orbital AM density* in the optical field [14,31,34,37,50,84]:

$$\boxed{\mathbf{L} = \mathbf{r} \times \mathbf{P}}. \tag{2.10}$$

This orbital AM density is an *extrinsic* and *transverse* quantity, akin to the mechanical AM (2.1)–(2.3). Furthermore, since it is defined via the momentum density $\mathbf{P}$, it is not an independent property of the field. Thus, the orbital AM (2.10) *locally* has properties (I) of the mechanical AM. However, as we show below, the *integral* orbital AM of the field, $\langle \mathbf{L} \rangle$, can exhibit quite different properties.

The energy, momentum and orbital AM densities can be equally defined for scalar fields. The specific vector character of the electromagnetic field manifests itself in the *spin AM*. The spin AM density is truly *intrinsic* and can be written as [21,30,32,34,37,50,84]

$$\mathbf{S} = \frac{g}{2} \text{Im}\left(\mathbf{E}^* \times \mathbf{E} + \mathbf{H}^* \times \mathbf{H}\right). \quad (2.11)$$

This quantity also has a clear interpretation. Namely, the spin density (2.11) is proportional to the local *ellipticity* of the field polarization, and it is directed along the normal to the polarization ellipse (assuming contributions from both the electric and magnetic fields). In particular, the spin density (2.11) agrees with Eqs. (2.4)–(2.6) for a circularly-polarized plane electromagnetic wave. Thus, the spin AM density $\mathbf{S}$ is an *independent* dynamical property of the field, which is related to the polarization degrees of freedom. Obviously, the spin AM density (2.11) is *intrinsic*, but its direction with respect to the wave momentum is not specified in the generic case. This hints that the spin density can have both longitudinal and transverse components in structured fields.

The total AM density of the field is a sum of the spin and orbital parts: $\mathbf{J} = \mathbf{S} + \mathbf{L}$. However, the spin and orbital AM manifest themselves in very different manners in local light-matter interactions [5–20] (see Section 2.4), so that they should be considered as independent physical properties, corresponding to different degrees of freedom [28–37].

In addition to the energy, momentum, and AM characteristics (2.8)–(2.11), there is one more fundamental quantity, which is less known. This is the optical *helicity* (sometimes called optical *chirality*) density [93,94], which has recently attracted considerable attention [34,86–89,95–105]. Although in quantum particle physics the helicity is determined by a product of the spin and momentum, in electromagnetism it is a locally-independent property, which is related to the so-called *dual symmetry* between the electric and magnetic field [34,88,95,97,102,105–109]. The helicity density of a monochromatic field reads

$$K = -g \, \text{Im}\left(\mathbf{E}^* \cdot \mathbf{H}\right). \quad (2.12)$$

The energy, momentum, spin, and helicity densities ($W$, $\mathbf{P}$, $\mathbf{S}$, and $K$) are independent quantities, and they form a complete set of fundamental dynamical properties of light important for our study. (In relativistic problems, one has to add the boost momentum related to the Lorentz transformations [33,34,110–113].)

Notably, the above dynamical characteristics (2.8)–(2.12) of classical optical fields are perfectly consistent with the *quantum-mechanical* approach. Namely, they represent the local expectation values of the corresponding first-quantization operators, when we introduce the wave function proportional to the properly-normalized electric and magnetic fields: $\psi = \sqrt{g/2}\begin{pmatrix} \mathbf{E} \\ \mathbf{H} \end{pmatrix}$ [30,34,50,89,114]. In doing so, equations (2.8)–(2.12) can be written as

$$W = \psi^\dagger \cdot (\omega) \psi, \quad \mathbf{P} = \text{Re}\left[\psi^\dagger \cdot (\hat{\mathbf{p}}) \psi\right], \quad \mathbf{L} = \psi^\dagger \cdot (\hat{\mathbf{L}}) \psi,$$

$$\mathbf{S} = \psi^\dagger \cdot (\hat{\mathbf{S}}) \psi, \quad K = \psi^\dagger \cdot (\hat{K}) \psi. \quad (2.13)$$

Here different operators act on different degrees of freedom of the wave function $\psi$. Specifically, the momentum $\hat{\mathbf{p}} = -i\nabla$ and orbital AM $\hat{\mathbf{L}} = \mathbf{r} \times \hat{\mathbf{p}}$ act on the spatial distribution of the wave function $\psi(\mathbf{r})$ (vector in Hilbert space), whereas the spin AM operator $\hat{\mathbf{S}}$, Eq. (2.5), acts on vector degrees of freedom in the complexified 3-dimensional space (it acts as $\mathbf{E}^* \cdot (\hat{\mathbf{S}}) \mathbf{E} = \text{Im}\left(\mathbf{E}^* \times \mathbf{E}\right)$ on vectors $\mathbf{E}$ and $\mathbf{H}$). In turn, the helicity operator $\hat{K} = -i\begin{pmatrix} 0 & -1 \\ 1 & 0 \end{pmatrix}$ (generator of the dual SO(2) transformation) mixes electric and magnetic degrees of freedom, i.e., it acts on the two-vector $\begin{pmatrix} \mathbf{E} \\ \mathbf{H} \end{pmatrix}$. In this formalism, the stationary free-space Maxwell equations can be written as $\hat{K}\psi = \left(\dfrac{\hat{\mathbf{S}} \cdot \hat{\mathbf{p}}}{k}\right)\psi$ [89,105,114], which proves the equivalence of the dual-rotation helicity $\hat{K}$ and the quantum-particle helicity $\hat{\mathbf{S}} \cdot \hat{\mathbf{p}}/k$ for Maxwell fields.

Alongside the *local* densities in Eqs. (2.8)–(2.12), we will also use the *integral* (mean) dynamical characteristics of the fields, denoted by angular brackets $\langle ... \rangle$. Usually, for quantum particles, this implies 3D integration over the whole space:

$$\langle W \rangle = \int W \, d^3\mathbf{r}, \quad \langle \mathbf{P} \rangle = \int P \, d^3\mathbf{r}, \quad \langle \mathbf{S} \rangle = \int S \, d^3\mathbf{r}, \quad \text{etc.} \tag{2.14}$$

However such integrals are well-defined only for localized fields, such as *wave packets*. Monochromatic fields (such as *optical beams*) cannot be localized in three dimensions, and the spatial integrals (2.14) diverge. Due to this, usually one uses two-dimensional integrals over the beam cross-section (i.e., the $(x,y)$-plane for $z$-propagating beams), which determine the linear densities of the corresponding quantities per unit $z$-length of the beam. (It also makes sense to determine the *fluxes* of the corresponding quantities through the beam cross-section [37,115].) Therefore, for monochromatic beams, we will imply integral properties obtained via the integration over the transverse cross-section plane:

$$\langle W \rangle = \int W \, d^2\mathbf{r}_\perp, \quad \langle \mathbf{P} \rangle = \int P \, d^2\mathbf{r}_\perp, \quad \langle \mathbf{S} \rangle = \int S \, d^2\mathbf{r}_\perp, \quad \text{etc.} \tag{2.14a}$$

Furthermore, for extended *periodic* field distributions (such as two-wave interference in Section 3.2), the spatial integration over one period will be implied.

### *2.3. Longitudinal spin and orbital AM in optical beams*

To illustrate the appearance of the above dynamical characteristics in an optical field, let us consider an axially-symmetric *paraxial polarized vortex beam* propagating in the $z$-direction. Its electric and magnetic fields can be written as

$$\mathbf{E} \simeq A(\rho,z) \frac{\overline{\mathbf{x}} + m\overline{\mathbf{y}}}{\sqrt{1+|m|^2}} \exp(ikz + i\ell\varphi), \quad \mathbf{H} \simeq \overline{\mathbf{z}} \times \mathbf{E}. \tag{2.15}$$

Here $m$ is the complex parameter, which characterizes the polarization [50,54,116], $\ell = 0, \pm 1, \pm 2, ...$ is the topological charge of the vortex, and $A(\rho,z)$ is the complex envelope amplitude of the beam. In the paraxial approximation (2.15), we neglect the small longitudinal $z$-components of the electric and magnetic fields (that is the approximate-equality symbol "$\simeq$" is used). These components become crucial for spin-orbit interaction and transverse-spin phenomena in non-paraxial fields.

The complex polarization parameter $m$ determines the real normalized Stokes parameters defined as ($\tau^2 + \chi^2 + \sigma^2 = 1$)

$$\tau = \frac{1-|m|^2}{1+|m|^2}, \quad \chi = \frac{2\operatorname{Re} m}{1+|m|^2}, \quad \sigma = \frac{2\operatorname{Im} m}{1+|m|^2}. \tag{2.16}$$

These three parameters describe the degrees of the $x/y$ linear polarizations, $45°/-45°$ linear polarizations, and right-hand/left-hand circular polarizations, respectively (the shapes of the Greek letters $\tau$, $\chi$, and $\sigma$ resemble horizontal/vertical, diagonal, and circular polarizations). The third Stokes parameter $\sigma$ is the *helicity parameter*, which takes on values $\pm 1$ for circularly-polarized waves, as in Eq. (2.4). Its direct relation to the helicity density (2.12) is revealed below.

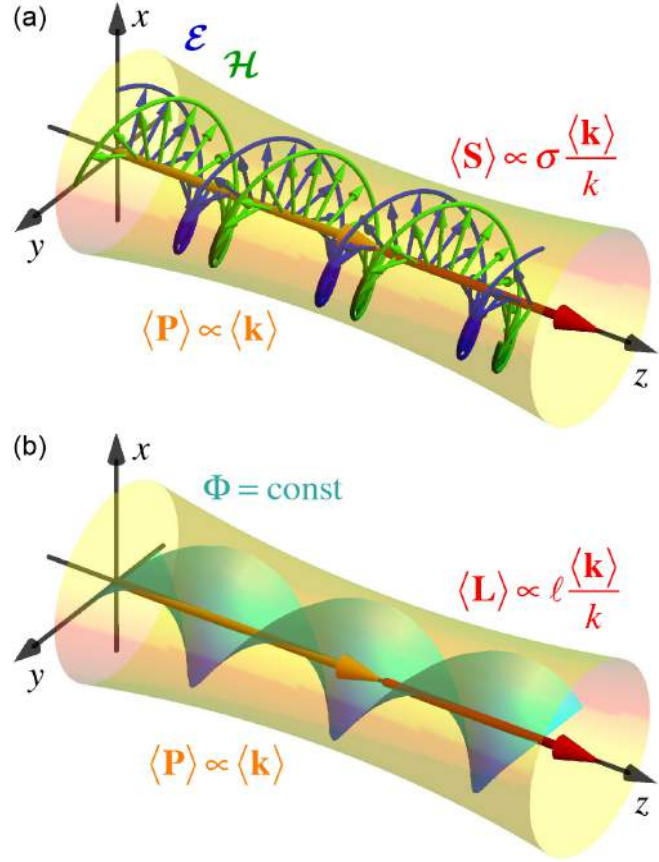

**Fig. 2.** Longitudinal spin and orbital angular momenta (2.18) and (III), (IV) in a paraxial optical beam (2.15) [1–14]. The beam carries integral momentum $\langle \mathbf{P} \rangle$ determined by its mean wave vector $\langle \mathbf{k} \rangle \simeq k\bar{\mathbf{z}}$. The spin AM $\langle \mathbf{S} \rangle$ is generated by the circular polarization and determined by its helicity parameter $\sigma$, whereas the orbital AM $\langle \mathbf{L} \rangle$ is produced by the helical phase, i.e., optical vortex of charge $\ell$. The instantaneous electric and magnetic fields $\mathcal{E}(z,0)$ and $\mathcal{H}(z,0)$ are shown in **(a)** for the right-hand circular polarization (parameter $m = i$, i.e., $\sigma = 1$). The constant-phase surface $\Phi = kz + \ell\varphi = 0$ is shown in **(b)** for the vortex with $\ell = 2$.

Substituting the field (2.15) into the general equations (2.8)–(2.12), we obtain the energy, momentum, spin AM, orbital AM, and helicity densities in the paraxial beam:

$$W \simeq g|A|^2 \omega, \quad \mathbf{P} \simeq \frac{W}{\omega}\left(k\bar{\mathbf{z}} + \frac{\ell}{\rho}\bar{\boldsymbol{\varphi}}\right), \quad \boxed{\mathbf{L} \simeq \frac{W}{\omega}(-\rho k\bar{\boldsymbol{\varphi}} + \ell\bar{\mathbf{z}})},$$

$$\boxed{\mathbf{S} \simeq \frac{W}{\omega}\sigma\bar{\mathbf{z}}}, \quad K \simeq \frac{W}{\omega}\sigma. \tag{2.17}$$

Equations (2.17) show a natural and intuitively clear picture of the beam properties, all proportional to the same intensity factor $g|A|^2$. First, we note that the beam has an energy density $W$ proportional to the frequency $\omega$ and the longitudinal momentum density $P_z$ proportional to the wave number $k = \omega/c$. Second, the beam carries the longitudinal spin AM density $\mathbf{S} = \sigma\bar{\mathbf{z}}$, similar to that for the plane wave, Eqs. (2.4)–(2.6) (Fig. 2a). The helicity density $K$ is naturally determined by the helicity parameter $\sigma$, and is in agreement with the longitudinal spin AM and momentum densities: $K = \mathbf{S} \cdot \mathbf{P}/P_z$. At the same time, the momentum has a transverse (with respect to the beam axis) *azimuthal* component $P_\varphi$ produced by the helical phase of the optical vortex, which generates the *longitudinal* orbital AM density $L_z$ proportional to $\ell$ [3–18,21] (Fig. 2b); this is a signature of vortex modes Eq. (2.7).

Integrating the densities (2.17) over the $(x,y)$ cross-section, we determine the *integral* values (2.14a) of the dynamical characteristics of the beam:

$$\langle W \rangle \propto \omega, \quad \langle \mathbf{P} \rangle \propto \langle \mathbf{k} \rangle, \quad \boxed{\langle \mathbf{L} \rangle \propto \ell \frac{\langle \mathbf{k} \rangle}{k}},$$

$$\boxed{\langle \mathbf{S} \rangle \propto \sigma \frac{\langle \mathbf{k} \rangle}{k}}, \quad \langle K \rangle \propto \sigma. \tag{2.18}$$

Here we introduced the mean wave vector, which is $\langle \mathbf{k} \rangle \simeq k \overline{\mathbf{z}}$ in the paraxial approximation, and wrote Eqs. (2.18) in the form valid for an arbitrary propagation direction of the beam.

The transition from the local densities (2.17) to the integral values (2.18) reveals two important peculiarities of the orbital AM in vortex states. First, the azimuthal momentum $P_\varphi$ and AM $L_\varphi$ disappear after the integration. As a result, the integral momentum $\langle \mathbf{P} \rangle$ and the orbital AM $\langle \mathbf{L} \rangle$ become *parallel* and *independent* properties! Second, the longitudinal orbital AM $\langle \mathbf{L} \rangle$ becomes *intrinsic*, i.e., independent of the choice of the coordinate origin: transformation (2.2) does not affect $\langle \mathbf{L} \rangle$ in Eq. (2.18) [21]. Thus, *the momentum-dependent extrinsic transverse local density $\mathbf{L} = \mathbf{r} \times \mathbf{P}$ produces an independent intrinsic longitudinal quantity* $\langle \mathbf{L} \rangle \| \langle \mathbf{P} \rangle$, Fig. 2. This amazing transmutation of the AM in the transition from its local to integral forms is a signature of structured fields (vortices in this case), which combine both wave and particle properties.

Summarizing the above properties of the integral spin and orbital AM of a paraxial beam:

**Paraxial beam AM:** Spin, Intrinsic, Longitudinal. Key parameters: $\sigma, \langle \mathbf{k} \rangle$. (III)

**Paraxial beam AM:** Orbital, Intrinsic, Longitudinal. Key parameters: $\ell, \langle \mathbf{k} \rangle$. (IV)

The vortex charge $\ell$ has the same $\mathcal{P}$-odd and $\mathcal{T}$-even features as the helicity $\sigma$, which ensures the proper $\mathcal{T}$-odd nature of the intrinsic orbital AM of a vortex beam.

Note that paraxial vortex beams (2.15) with circular polarizations ($m = \pm i$, $\sigma = \pm 1$) are approximate eigenmodes of the operators $\hat{p}_z$, $\hat{L}_z$, $\hat{S}_z$, and $\hat{K}$ with the corresponding eigenvalues $k$, $\ell$, $\sigma$, and $\sigma$. Such states provide a convenient basis for performing operations with paraxial photons carrying well-defined momentum, spin, and orbital AM in quantum optics [11–13].

The proportionality of the spin and orbital AM to the polarization helicity $\sigma$ and vortex charge $\ell$ holds true only in the paraxial approximation. Taking into account the longitudinal field components $E_z$ and $H_z$ in *non-paraxial* beams results in *spin-to-orbital AM conversion* [19,20,31,37,117–123]. This effect originates from the helicity-dependent vortex structure $\exp(i\sigma\varphi)$ of the longitudinal components of a circularly-polarized beam field, which, in turn, appears due to the transversality of the Maxwell fields. Introducing the typical convergence angle $\vartheta$ of a non-paraxial (focused) vortex beam as $\cos\vartheta = c \langle P_z \rangle / \langle W \rangle$, the integral spin and orbital AM in such beam take the form [31,35,37,121,123]

$$\langle \mathbf{S} \rangle \propto \sigma \frac{\langle \mathbf{k} \rangle}{\langle k \rangle} \cos\vartheta \;, \quad \langle \mathbf{L} \rangle \propto \left[ \ell + \sigma(1-\cos\vartheta) \right] \frac{\langle \mathbf{k} \rangle}{\langle k \rangle} \;. \tag{2.19}$$

Here we denoted $\langle k \rangle \equiv |\langle \mathbf{k} \rangle|$ and implied that the beam is constructed as a superposition of circularly-polarized plane waves with $\sigma = \pm 1$, and, hence, possesses a well-defined helicity $\langle K \rangle \propto \sigma$. Equations (2.19) indicate that changing the paraxiality (focusing the beam) does not change its total AM $\langle \mathbf{J} \rangle = \langle \mathbf{S} \rangle + \langle \mathbf{L} \rangle$ but results in a redistribution between the spin and orbital parts. Notably, this spin-to-orbital AM conversion can achieve 100% efficiency for $\vartheta = \pi/2$ [124]. For a detailed analysis and overview of the spin-orbit interactions in non-paraxial fields we refer the reader to [31,37,123].

We emphasize that the spin and orbital AM (2.19) follow from the same Eqs. (2.9)–(2.11), which are valid for any monochromatic fields. Furthermore, the canonical expressions (2.9)–(2.11) do not contain the extra surface term, which affects non-paraxial calculations with the Poynting momentum [26,27,35].

Focusing optical beams reveals one more feature of the wave-particle duality. It can be shown from relativistic considerations that any object carrying intrinsic AM cannot be shrunk to a point and its minimum size is proportional to its intrinsic AM [112]. The same conclusion occurs for the minimum size of a tightly-focused beam carrying intrinsic spin and orbital AM [125]. Thus, despite the fact that the integral intrinsic AM is well-defined for optical beams or wave packets, the finite spatial extension is crucial for it. This also follows from the non-localizability of photons with spin and the non-commutative character of their covariant coordinates [31,126,127].

### *2.4. Local measurements of the spin and orbital AM*

After we described the main properties of the momentum and angular momenta of free optical fields, it is important to discuss how these properties can be measured experimentally. Since the *integral* spin and orbital AM (2.18) in paraxial beams look very similar, they manifest themselves similarly in experiments with the detector size larger than the beam cross-section [2,4,11–13,128]. In contrast, *local* measurements of the spin and orbital AM reveal their *densities* (2.10) and (2.11) or (2.17), which have very different properties [16–20,29,34,37,50,92]. This was first demonstrated in remarkable experiments [16,17], where small probe particles were employed to reveal the local AM properties in paraxial vortex beams, Fig. 3. It turned out that such a probe particle *spins* around its axis, proportionally to the local spin AM density $S_z$ (circular polarization) of the field, and also *orbits* around the vortex-beam axis proportionally to its orbital AM. The latter motion can be considered as the local translational motion proportional to the azimuthal canonical momentum $P_\varphi$ (2.17) in the vortex beam. Thus, the spin and orbital degrees of freedom manifest themselves differently in local interactions with matter and can be separately measured in optical fields. Similar experiments with probe particles were also employed to detect the spin-to-orbital AM conversion in [19,20].

To understand the results of the experiments [16–20], we now consider the interaction of a small isotropic spherical particle with a monochromatic optical field. A typical subwavelength (Rayleigh) particle interacts with the field via the *electric-dipole* coupling. The particle is characterized by its complex electric polarizability $\alpha^e$, which generates a complex electric dipole moment $\boldsymbol{\pi} = \alpha^e \mathbf{E}$ induced by the electric wave field. Although here we consider the dipole coupling with a *classical* particle, similar results take place in quantum interactions with *atoms or molecules* [24,29,79,129–131].

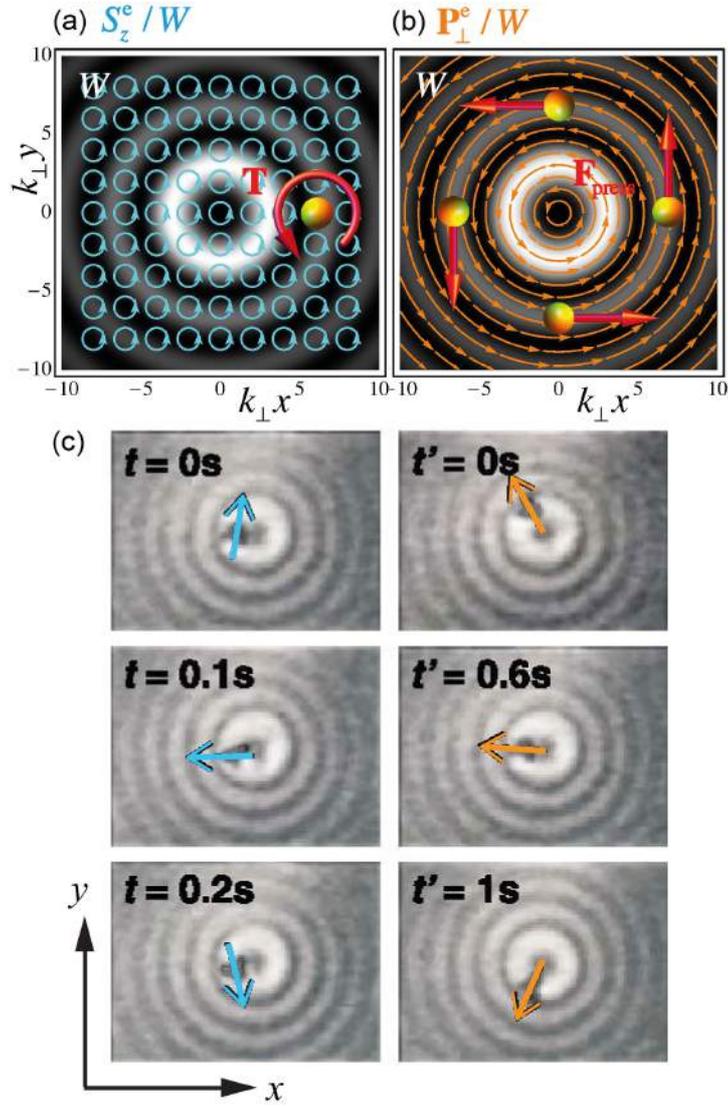

**Fig. 3.** Local spin AM and transverse-momentum densities and their measurements in the paraxial Bessel beam given by Eq. (2.15) with $A(\rho,z) = J_{|\ell|}(k_\perp \rho)$ and $k \to k_z$, where $k_\perp = \sqrt{k^2 - k_z^2} \ll k$. The right-hand circularly polarized beam ($m = i, \sigma = 1$) with a vortex of charge $\ell = 2$ is shown via its energy-density distributions $W(r)$ (grayscale plots). **(a)** The distribution of the transverse circular polarization of the electric field [in cyan] indicates the longitudinal spin AM density $S_z^e$, Eqs. (2.11) and (2.17). **(b)** The distribution of the transverse (azimuthal) canonical momentum density $\mathbf{P}_\perp^e$ [in orange] from the helical phase (optical vortex), Eqs. (2.9) and (2.17), produces the corresponding orbital AM density $L_z^e = r P_\varphi^e$. **(c)** Experimental measurements from [17]. The spin AM and canonical momentum densities immediately reveal themselves when interacting with a small probe particle. The particle spins due to a torque proportional to the spin density $S_z^e$ and orbits (i.e., moves azimuthally) due to the radiation-pressure force proportional to the momentum density $P_\varphi^e$, Eqs. (2.22) and (2.23). The radial trapping of the particle at the intensity maximum occurs due to the gradient force in Eq. (2.22). These measurements clearly show the different intrinsic and extrinsic nature of the spin and orbital AM densities and their separate observability.

It is important to note that most of the natural particles interact very weakly with the magnetic field $\mathbf{H}$ via the induced magnetic dipole moment $\boldsymbol{\mu} = \alpha^m \mathbf{H}$, i.e., have a very small magnetic polarizability $\alpha^m$: $|\alpha^m| \ll |\alpha^e|$. This *breaks the dual (electric-magnetic) symmetry* of the free electromagnetic field [105] and makes the probe sensitive to the *electric* parts of all quantities [34,89]. Indeed, all the dynamical field properties (2.8)–(2.11), but not the helicity (2.12), naturally represent sums of the electric and magnetic contributions:

$$W = W^e + W^m, \quad \mathbf{P} = \mathbf{P}^e + \mathbf{P}^m, \quad \mathbf{L} = \mathbf{L}^e + \mathbf{L}^m, \quad \mathbf{S} = \mathbf{S}^e + \mathbf{S}^m. \quad (2.20)$$

Most of the standard laboratory measurements reveal only the electric parts of these quantities. The electric and magnetic contributions in Eqs. (2.20) are equivalent in paraxial optical fields, including the beams (2.15)–(2.18), but can differ significantly in generic structured fields [30,50,54]. It is worth noticing that definitions of the momentum, spin and orbital AM densities depend on the choice of the Lagrangian of the free electromagnetic field [34]. For instance, these are "electrically-biased", $\mathbf{P} \to 2\mathbf{P}^e$, $\mathbf{S} \to 2\mathbf{S}^e$ and $\mathbf{L} \to 2\mathbf{L}^e$, with the standard dual-asymmetric Lagrangian density $\mathcal{L} \propto (\mathcal{E}^2 - \mathcal{H}^2)$ [34,37], while the dual-symmetric definitions (2.9)–(2.11) are the most natural choice for a free field [30,32,34,37].

The energy, momentum, and angular-momentum transfer in light-particle interactions can be quantified by the *absorption rate* A, *optical force* **F**, and *torque* **T** on the particle, respectively. Calculating these quantities in the electric dipole-coupling approximation ($\alpha^m \simeq 0$), one can derive [50,89,132–134]:

$$A = g^{-1} \text{Im}(\alpha^e) W^e, \quad (2.21)$$

$$\mathbf{F} = g^{-1} \left[ \frac{1}{2\omega} \text{Re}(\alpha^e) \nabla W^e + \text{Im}(\alpha^e) \mathbf{P}^e \right] \equiv \mathbf{F}_{\text{grad}} + \mathbf{F}_{\text{press}}, \quad (2.22)$$

$$\mathbf{T} = g^{-1} \text{Im}(\alpha^e) \mathbf{S}^e. \quad (2.23)$$

The first and second terms in Eq. (2.22) represent the gradient and radiation-pressure optical forces, respectively. Equations (2.21)–(2.23) show that the absorption rate, radiation-pressure force, and torque on the particle "measure" the energy, canonical-momentum, and spin-AM densities in the wave field. In particular, Eqs. (2.22) and (2.23) with Eqs. (2.17) perfectly explain the results of the experiments [16,17] (Fig. 3). Namely: (i) the probe particle spins about its center proportionally to the spin AM density $S_z^e$, (ii) it is trapped in the annular field intensity maximum due to the radial gradient force proportional to $\partial W^e / \partial r$, and (iii) it orbits along the circular trajectory due to the radiation-pressure force produced by the azimuthal canonical momentum $P_\varphi^e$. Importantly, in local measurements, the orbital AM does not represent an independent degree of freedom, but it is produced by the canonical momentum density according to Eq. (2.10). Thus, local measurements reveal the intrinsic nature of the spin AM and locally-extrinsic nature of the orbital AM. Basic optical forces and torques (2.22) and (2.23) play a key role in numerous problems and applications using mechanical properties of light, such as optical trapping and manipulation of atoms, molecules, and small particles, as well as in various optomechanical phenomena (see [135–139] for reviews).

It should be emphasized that Eqs. (2.22) and (2.23) support the *canonical* picture of the optical momentum and AM, Eqs. (2.9)–(2.11), rather than the kinetic one based on the *Poynting vector* [34,37,38,50,54,84]. Indeed, the local momentum transfer in light-matter interactions is determined by the canonical momentum (2.9). For example, the "supermomentum" transfer larger than $k = \omega/c$ per photon can be observed near optical-vortex cores and in evanescent waves with "superluminal" canonical momentum $|\mathbf{P}|/W > c$ [30,50,92,131,140,141], while the Poynting vector is always "subluminal": $|\mathbf{\Pi}|/W < c$. Moreover, careful considerations show that a number of other (completely different) methods of measurements of the local momentum density in optical fields [142–147] results in the same canonical momentum **P** (or $\mathbf{P}^e$) rather than the Poynting vector. In particular, measurements of the azimuthal momentum in optical beams reveal the canonical momentum $P_\varphi$ (2.17) due to the optical vortex [142–146], but they cannot measure the azimuthal Poynting vector $\Pi_\varphi$ caused by the circular polarization [14]. In addition, the so-called *quantum weak measurements* of the photon momentum density immediately yield the canonical momentum owing to its intimate relation to the momentum operator $\hat{\mathbf{p}}$ [30,34,92,147].

The helicity density (2.12) does not manifest itself in interaction with spherical particles, but it becomes crucial in the interaction with *chiral* particles. The chirality of the particle or a molecule mixes the electric and magnetic interactions, so that the chiral contributions to the induced electric and magnetic dipole moments can be written as [85,86,89,103]

$$\boldsymbol{\pi}^{\text{ch}} = -i\alpha^{\text{ch}}\mathbf{H}, \quad \boldsymbol{\mu}^{\text{ch}} = i\alpha^{\text{ch}}\mathbf{E}, \qquad (2.24)$$

where $\alpha^{\text{ch}}$ is the chiral polarizability of the particle. Recently, it was noticed that the absorption-rate term caused by the chirality of the particle is proportional to the local *helicity density* (2.12) of the field [86,89]:

$$\text{A}^{\text{ch}} = -g^{-1}\omega \,\text{Im}\left(\alpha^{\text{ch}}\right) K. \qquad (2.25)$$

In combination with the natural dual asymmetry of molecules, this has resulted in a new method for the enhancement of the optical circular dichroism and enhanced optical enantioselectivity [86,98–101,148,149].

While the natural circular dichroism due to the particle chirality is coupled to the helicity density of the field, the *magnetic circular dichroism* (MCD), induced by the presence of an external constant magnetic field $\mathbf{H}_0$, turns out to be coupled to the *spin AM density* (2.11) [150,151]. According to [150] the $\mathbf{H}_0$-induced term in the electric dipole polarizability of an isotropic particle can be written as

$$\boldsymbol{\pi}^{\text{MCD}} = i\alpha^{\text{MCD}}\left(\mathbf{E}\times\mathbf{H}_0\right). \qquad (2.26)$$

Calculating the corresponding MCD contribution to the absorption rate, $\text{A}^{\text{MCD}} = \frac{\omega}{2}\text{Im}\left(\boldsymbol{\pi}^{\text{MCD}}\cdot\mathbf{E}^*\right)$, and using Eqs. (2.11) and (2.20), we arrive at

$$\text{A}^{\text{MCD}} = -g^{-1}\omega\,\text{Im}\left(\alpha^{\text{MCD}}\right)\left(\mathbf{S}^{\text{e}}\cdot\mathbf{H}_0\right). \qquad (2.27)$$

A similar coupling of the electric spin density to an external magnetic field also occurs in atomic transitions in light-atom interactions [152–154].

Thus, the natural circular dichroism and MCD provide efficient methods of non-mechanical local measurements of the helicity and spin AM densities in optical fields. Note that such correspondence looks natural from spatial-inversion and time-reversal symmetry arguments. Indeed, the chirality and helicity are both $\mathcal{P}$-odd and $\mathcal{T}$-even scalar properties, while the magnetic field and spin AM are both $\mathcal{P}$-even and $\mathcal{T}$-odd vector quantities [85,89].

## 3. Transverse spin angular momenta

In this Section we consider a new kind of angular momentum of light: the *transverse spin AM*. This angular momentum exhibit features which are in sharp contrast to the usual longitudinal spin AM in Eqs. (2.6), (2.17), (2.18), and to what is known about the spin of photons. As we show below, there are different types of transverse spin, which can have properties dramatically different from (II) and (III). Throughout this Section we do not consider the orbital AM (2.10) because in all examples here it does not represent an independent degree of freedom.

*3.1. Evanescent waves*

3.1.1. Transverse (out-of-plane) helicity-independent spin. The first example of the transverse spin in *evanescent waves* was described by Bliokh and Nori [46,50] and a bit later by Kim *et al*. [48,59]. A single evanescent wave is one of the simplest solutions of free-space Maxwell equations [155]. Assuming that the wave propagates along the $z$-axis and decays in the $x > 0$ half-space, it can be represented as a plane wave with the *complex* wave vector

$$\mathbf{k} = k_z \bar{\mathbf{z}} + i\kappa \bar{\mathbf{x}}, \quad k^2 = k_z^2 - \kappa^2. \tag{3.1}$$

The electric and magnetic fields of the evanescent wave with an arbitrary polarization can be written as [50]

$$\mathbf{E} = \frac{A_0}{\sqrt{1+|m|^2}}\left(\bar{\mathbf{x}} + m\frac{k}{k_z}\bar{\mathbf{y}} - i\frac{\kappa}{k_z}\bar{\mathbf{z}}\right)\exp(ik_z z - \kappa x), \quad \mathbf{H} = \frac{\mathbf{k}}{k}\times\mathbf{E}, \tag{3.2}$$

where $A_0$ is a constant field amplitude.

Even though the electric wave field (3.2) has all three components, its polarization degrees of freedom are described by the same complex polarization parameter $m$ and the corresponding Stokes parameters as in Eqs. (2.15) and (2.16). Therefore, we will call the $m = 0$ and $m = \infty$ waves with $\tau = \pm 1$ as "linearly polarized", the $m = \pm 1$ waves with $\chi = \pm 1$ as "diagonally polarized", and the $m = \pm i$ waves with $\sigma = \pm 1$ as "circularly polarized", even though the actual 3D polarization vector can be rather nontrivial [50]. The evanescent wave (3.2) propagates along the $z$-axis with "anomalous" wave momentum $k_z > k$, and it decays exponentially in the positive $x$-direction. Such waves can be generated, e.g., in total internal reflection at the glass-air interface $x = 0$ [155], as shown in Fig. 4. The linearly-polarized evanescent waves with $m = 0$ (TM mode) or $m = \infty$ (TE mode) are also parts of surface electromagnetic waves at the $x = 0$ interface between two optical media [56,156]: e.g., the surface plasmon-polaritons at the metal-air interface [157]. Thus, the wave (3.2) is physically meaningful and well-defined only in the *half*-space $x > 0$.

Most importantly, the imaginary component of the wave vector (3.1) and the plane-wave transversality condition $\nabla\cdot\mathbf{E} = \mathbf{k}\cdot\mathbf{E} = 0$ result in the *"imaginary" longitudinal component* $E_z \propto -i\kappa/k_z$ in the field (3.2). This component, together with the "real" component $E_x$, produces a rotation of the electric field (i.e., effective *elliptical* polarization) in the $(x,z)$ plane of the wave propagation [46,50]. Since the field also propagates along the $z$-axis, the electric field follows a *cycloidal* trajectory, like a point on a moving and spinning wheel, see Fig. 4 [49,50,158]. This is in contrast to the *helical* trajectory in the case of the usual circular polarization and the longitudinal spin, Figs. 1b and 2a. Nonetheless, according to Eq. (2.11), the elliptical polarization in the $(x,z)$-plane implies the spin AM density directed along the transverse $y$-axis, i.e., *orthogonally* to the momentum and wave vector of the evanescent field, Figs. 4 and 5.

To quantify the main properties of the evanescent wave, we substitute the field (3.2) into the general equations (2.8)–(2.12). Thus, we obtain the energy, momentum, spin AM, and helicity densities:

$$W = g|A_0|^2 \omega \exp(-2\kappa x), \quad \mathbf{P} = \frac{W}{\omega}k_z\bar{\mathbf{z}}, \tag{3.3}$$

$$\mathbf{S} = \frac{W}{\omega}\left(\sigma \frac{k}{k_z}\overline{\mathbf{z}} + \boxed{\frac{\kappa}{k_z}\overline{\mathbf{y}}}\right), \qquad K = \frac{W}{\omega}\sigma. \tag{3.4}$$

The energy and momentum densities (3.3) show intuitively-clear values proportional to $\omega$ and $\text{Re}\,\mathbf{k} = k_z\overline{\mathbf{z}}$, and decaying with $x$ together with the wave intensity, Fig. 5b. The "superluminal" canonical momentum (3.3) with $k_z > \omega/c$ is in contrast to the "subluminal" Poynting vector [46,92], and it was measured in light-atom interactions in [140,141]. The spin AM density in Eq. (3.4) contains the usual longitudinal component $S_z$, which is similar to the spin AM of propagating waves (2.6) and (2.17). It is directed along the wave momentum $\text{Re}\,\mathbf{k}$ and is proportional to the helicity parameter $\sigma$. In addition, the helicity density $K$ in Eq. (3.4) is also similar to that in a propagating plane wave, Eq. (2.17).

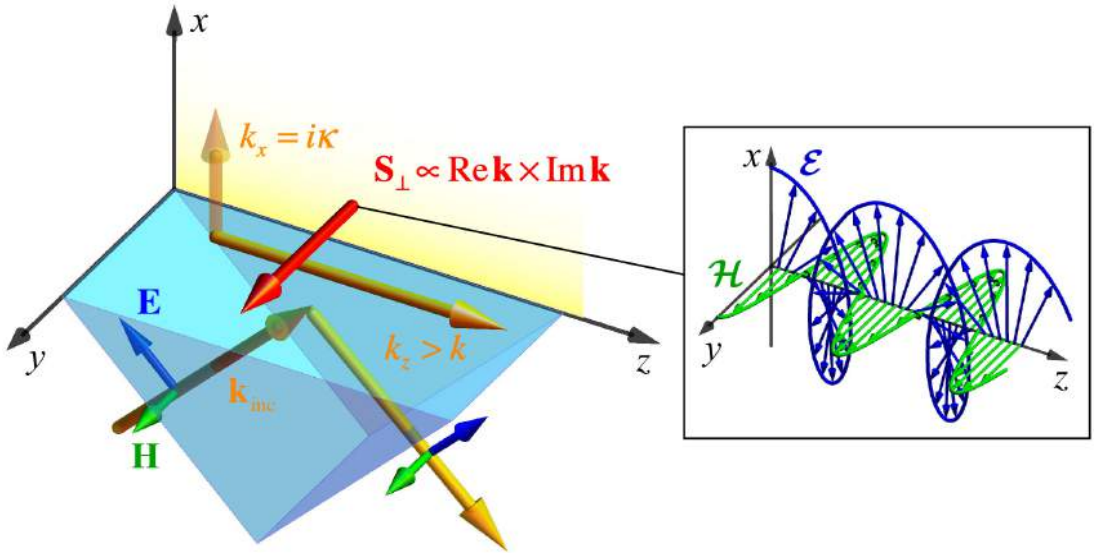

**Fig. 4.** Evanescent wave (3.2) generated by the total internal reflection of a plane wave and the transverse (out-of-plane) spin AM density (3.5) and (V) in the evanescent field [46,50]. The evanescent wave is characterized by the complex wave vector (3.1) $\mathbf{k}$ and is defined in the $x > 0$ half-space. The inset shows the $z$-evolution of the instantaneous electric and magnetic fields $\mathcal{E}(\mathbf{r},0)$ and $\mathcal{H}(\mathbf{r},0)$ for the simplest linear $x$-polarization (TM mode with $m = 0$, $\tau = 1$). The cycloidal rotation of the electric field within the propagation $(x,z)$-plane generates the helicity-independent transverse spin AM $\mathbf{S}_\perp$, Eqs. (3.4)–(3.7).

The only extraordinary term in the above equations is the *transverse spin AM density* $S_y$ (shown in the red frame), Figs. 4 and 5a. We write it separately in the form valid for an arbitrary propagation direction of the evanescent wave:

$$\boxed{\mathbf{S}_\perp = \frac{W}{\omega}\frac{\text{Re}\,\mathbf{k}\times\text{Im}\,\mathbf{k}}{(\text{Re}\,\mathbf{k})^2}}. \tag{3.5}$$

In contrast to the usual longitudinal spin considered in Section 2, the transverse spin (3.5) is completely *independent of the helicity* and other polarization parameters. Instead, it is solely determined by the *complex wave vector* of the evanescent wave and is directed *out of the plane* formed by its real and imaginary parts. Obviously, as any spin AM, it has an *intrinsic* nature. Importantly, the imaginary part of the wave vector, $\text{Im}\,\mathbf{k}$, is a $\mathcal{P}$-odd but $\mathcal{T}$-even quantity (because the time-reversal transformation implies complex conjugation for all complex quantities [89,159]), so that the transverse spin (3.5) has the proper $\mathcal{T}$-odd and $\mathcal{P}$-even symmetry. Summarizing the unusual properties of the transverse spin AM (3.5):

<div style="border: 1px solid red; padding: 8px;">
**Evanescent wave:** Spin, Transverse (out-of-plane), Key parameters: $\text{Re}\,\mathbf{k}, \text{Im}\,\mathbf{k}$. (V)
</div>

Remarkably, the crucial dependence of the transverse spin (3.5) on the wave-vector (momentum) is intimately related to the fundamental topological properties of Maxwell equations [56], and it has been employed for transverse spin-direction interfaces with evanescent waves [52,53,57,74–83] (see Section 3.3.5 below). According to the general Eqs. (2.22) and (2.23), the longitudinal momentum and transverse spin AM in evanescent waves can be directly measured via the local interaction with a probe particle. In particular, the transverse spin in Eq. (3.4) exerts the corresponding helicity-independent torque $T_y \propto S_y^e$ [50,51]. This is schematically shown in Fig. 5, and in Section 3.3 we consider the main experiments involving the transverse spin.

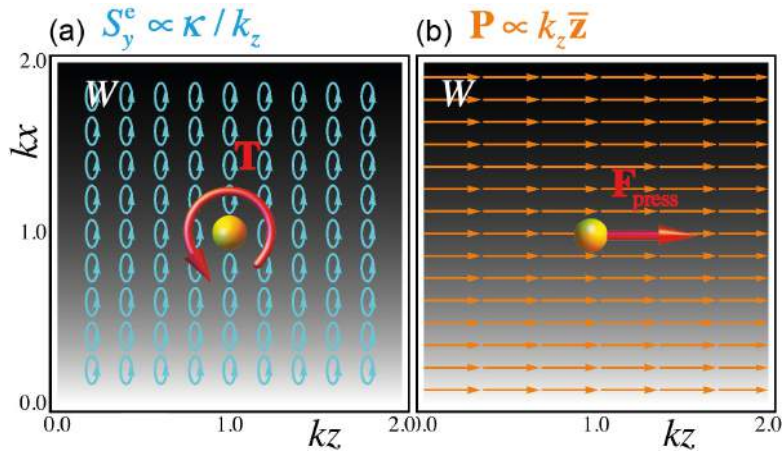

**Fig. 5.** The energy, momentum, and transverse spin AM densities (3.3)–(3.7) in the evanescent wave (3.2) with $k_z/k = 1.1$ and linear TM polarization ($m = 0$, $\tau = 1$, see Fig. 4). The energy density $W(x)$ is shown as the grayscale background distribution. **(a)** Akin to Fig. 3, the transverse spin AM density (3.6) $S_y^e$ [normalized by $W(x)$] is shown in the form of a polarization-ellipse distribution in the propagation $(x,z)$-plane. It exerts the transverse optical torque (2.23) $T_y$ on a probe particle [50,51,162]. **(b)** The longitudinal momentum density is naturally proportional to the wave number $k_z$, and it exerts the radiation-pressure force (2.22) $F_z$.

The fact that the transverse spin AM of evanescent waves was not discussed before 2012 is a remarkable example of preconception and the lack of communication between different physical communities. On the one hand, the imaginary longitudinal field component and the $(x,z)$-plane rotation of the electric field in evanescent waves were known for decades in optics and plasmonics (see, e.g., [155,157,160]). Furthermore, the in-plane elliptical polarization was measured in experiments [153,161]. On the other hand, knowing the direct connection between the elliptical polarization and spin AM since Poynting and Beth [1,2], the optical AM community has never considered *linearly*-polarized evanescent waves (e.g., $m = 0$ TM-modes in the case of surface plasmon-polaritons) as possible candidates for spin AM studies. There were even numerical simulations showing the transverse helicity-independent torque on a probe particle in the evanescent field [162], but this torque was not connected to the effective elliptical in-plane polarization, and the key conclusion about the presence of the transverse spin AM was not made until [46,50].

3.1.2. *Transverse (in-plane) dual-antisymmetric spin.* In contrast to the usual longitudinal spin AM in circularly-polarized paraxial waves, the transverse spin has *asymmetric electric and magnetic properties*. For instance, in the TM-polarized wave (3.2) with $m=0$, $\tau=1$, only the *electric* field rotates in the $(x,z)$-plane, while the magnetic field has only the $y$-component, as shown in Fig. 4. In this case, the transverse spin AM has a purely *electric* origin: $\mathbf{S}=\mathbf{S}^e$, $\mathbf{S}^m=0$, see Eqs. (2.11) and (2.20). If the wave has a linear $y$ (TE) polarization with $m=\infty$, $\tau=-1$, then the situation becomes the opposite: the *magnetic* field rotates in the $(x,z)$-plane, the electric field has only a $y$-component, and the transverse spin is purely *magnetic*: $\mathbf{S}=\mathbf{S}^m$, $\mathbf{S}^e=0$ [50]. The latter feature was mistakenly interpreted in [48] as the absence of the transverse spin AM in the TE-mode.

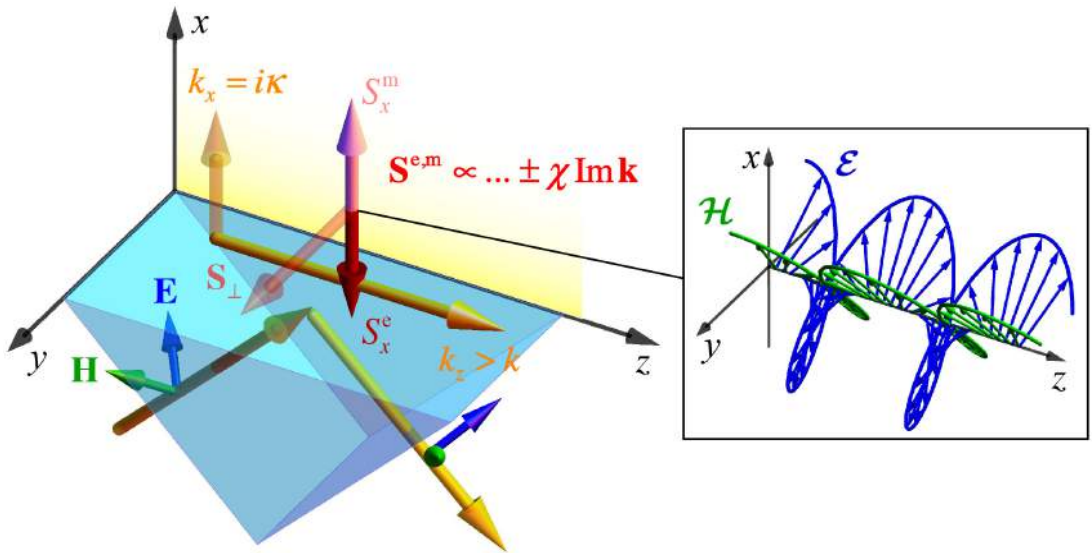

**Fig. 6.** The dual-antisymmetric transverse (in-plane) spin AM, Eqs. (3.6) and (3.7), in a diagonally-polarized evanescent wave (3.2) [50], cf., Fig. 4. A diagonal polarization with $m=-1$ and $\chi=-1$ is shown here, and the inset displays the $z$-evolution of the instantaneous electric and magnetic fields $\mathcal{E}(\mathbf{r},0)$ and $\mathcal{H}(\mathbf{r},0)$. In addition to the transverse (out-of-plane) spin AM $S_y$ shown in Fig. 4, the diagonal polarization produces $\chi$-dependent anti-parallel vertical (in-plane) electric and magnetic spin AM $S_x^e=-S_x^m$, Eqs. (3.6), (3.7), and (VI). These oppositely-directed electric and magnetic spins are generated by opposite $(y,z)$-plane rotations of the electric and magnetic fields in a diagonally-polarized evanescent wave.

For a generic polarization, the electric and magnetic parts of the spin AM density (2.11) and (2.20) in the evanescent wave (3.2) are [50]:

$$\mathbf{S}^{e,m} = \frac{W}{2\omega}\left(\sigma \frac{k}{k_z}\overline{\mathbf{z}} + (1\pm\tau)\frac{\kappa}{k_z}\overline{\mathbf{y}} \pm \chi \frac{\kappa k}{k_z^2}\overline{\mathbf{x}}\right), \qquad (3.6)$$

Here the transverse components $S_y^{e,m}$ show the above electric-magnetic asymmetry, which is controlled by the first Stokes parameter $\tau$. Moreover, Eq. (3.6) reveals a new, *vertical* term $S_x^{e,m}$ (shown in the green frame), which is directed along the wave *inhomogeneity*, $\text{Im}\,\mathbf{k}$, see Fig. 6. It is also *transverse* with respect to the wave propagation direction $\text{Re}\,\mathbf{k}$, but lies *in plane* with the complex wave vector $\mathbf{k}$. This *"dual-antisymmetric transverse (in-plane) spin AM"* has the opposite electric and magnetic contributions: $S_x^e = -S_x^m$, so that $S_x = 0$ [50]. Nonetheless, the vertical electric spin $S_x^e$ naturally contributes to light-matter interactions sensitive to the *electric* rather than magnetic parts of the fundamental quantities, see Section 2.4. The vertical spin AM $S_x^{e,m}$ is caused by the opposite rotations of the electric and magnetic fields projected on the $(y,z)$ plane for the waves polarized *diagonally* at $\pm 45°$ (Fig. 6). Therefore, it is proportional to the second Stokes parameter $\chi$.

Representing Eq. (3.6) in the general vector form based on the wave vector (3.1) yields

$$\mathbf{S}^{e,m} = \frac{W}{2\omega}\left(\sigma\frac{k\,\mathrm{Re}\,\mathbf{k}}{(\mathrm{Re}\,\mathbf{k})^2} + (1\pm\tau)\frac{\mathrm{Re}\,\mathbf{k}\times\mathrm{Im}\,\mathbf{k}}{(\mathrm{Re}\,\mathbf{k})^2} \pm \chi\frac{k\,\mathrm{Im}\,\mathbf{k}}{(\mathrm{Re}\,\mathbf{k})^2}\right). \qquad (3.7)$$

Here all the three terms must have the same $\mathcal{T}$-odd and $\mathcal{P}$-even character of the AM. Therefore, we conclude that the Stokes parameter $\tau$ is a $\mathcal{P}$-even and $\mathcal{T}$-even quantity, while the Stokes parameter $\chi$ is a $\mathcal{P}$-*odd* and $\mathcal{T}$-*odd* quantity. Summarizing the properties of the "dual-antisymmetric transverse spin AM" in Eq. (3.7):

> **Evanescent wave AM:** Spin, Transverse (in-plane), Anti-dual. Key parameters: $\chi$, $\mathrm{Im}\,\mathbf{k}$. (VI)

In this Subsection we analyzed only *local* momentum and AM densities. The reason for this is that the evanescent wave itself is defined only in a half-space. Since the wave (3.2) is homogeneous in the $y$ and $z$ directions, it only makes sense to calculate integral quantities (2.14) with the *integration over the $x > 0$ semi-axis*. This integration preserves all features of the $x$-dependent densities and can be symbolically written as

$$\langle W\rangle^+ = \frac{1}{2\kappa}W(0),\quad \langle\mathbf{P}\rangle^+ = \frac{1}{2\kappa}\mathbf{P}(0),\quad \langle\mathbf{S}\rangle^+ = \frac{1}{2\kappa}\mathbf{S}(0),\quad \langle K\rangle^+ = \frac{1}{2\kappa}K(0),\ \text{etc.} \qquad (3.8)$$

Here the "+" superscript stands for the positive-$x$ integration and the "0" arguments indicate the values at $x = 0$.

### *3.2. Propagating waves*

The transverse spin AM is not an exclusive feature of evanescent waves. A very similar transverse spin AM density also appears in basic configurations with freely propagating fields. The necessary condition of its appearance is the transverse *inhomogeneity* of the field intensity: in all cases the transverse-spin density is accompanied by transverse intensity gradients. It is these gradients that generate the "imaginary" longitudinal field components.

<u>3.2.1. Two-wave interference.</u> The simplest propagating-wave configuration with inhomogeneity is the *interference of two plane waves* propagating at an angle $2\gamma$ between their wave vectors. Below, we basically follow the analysis of the recent work [54]. It is convenient to choose the $z$-axis along the mean direction of propagation, and the $(x,z)$-plane as the *propagation plane* formed by the two wave vectors (see Fig. 7):

$$\mathbf{k}_{1,2} = k\cos\gamma\,\overline{\mathbf{z}} \pm k\sin\gamma\,\overline{\mathbf{x}} \equiv k_z\overline{\mathbf{z}} \pm k_x\overline{\mathbf{x}}. \qquad (3.9)$$

Then, the electric and magnetic fields of the two plane waves can be written as [cf., Eqs. (2.4) and (2.15)]

$$\mathbf{E}_{1,2} = \frac{A_0}{\sqrt{1+|m_{1,2}|^2}} \left( \frac{k_z}{k}\bar{\mathbf{x}} + m_{1,2}\bar{\mathbf{y}} \mp \frac{k_x}{k}\bar{\mathbf{z}} \right) \exp(i\Phi_{1,2}), \quad \mathbf{H}_{1,2} = \frac{\mathbf{k}_{1,2}}{k} \times \mathbf{E}_{1,2}, \tag{3.10}$$

where $\Phi_{1,2} = k_z z \pm k_x x$ are the wave phases, and we assume that the two waves have equal electric-field amplitudes $A_0$.

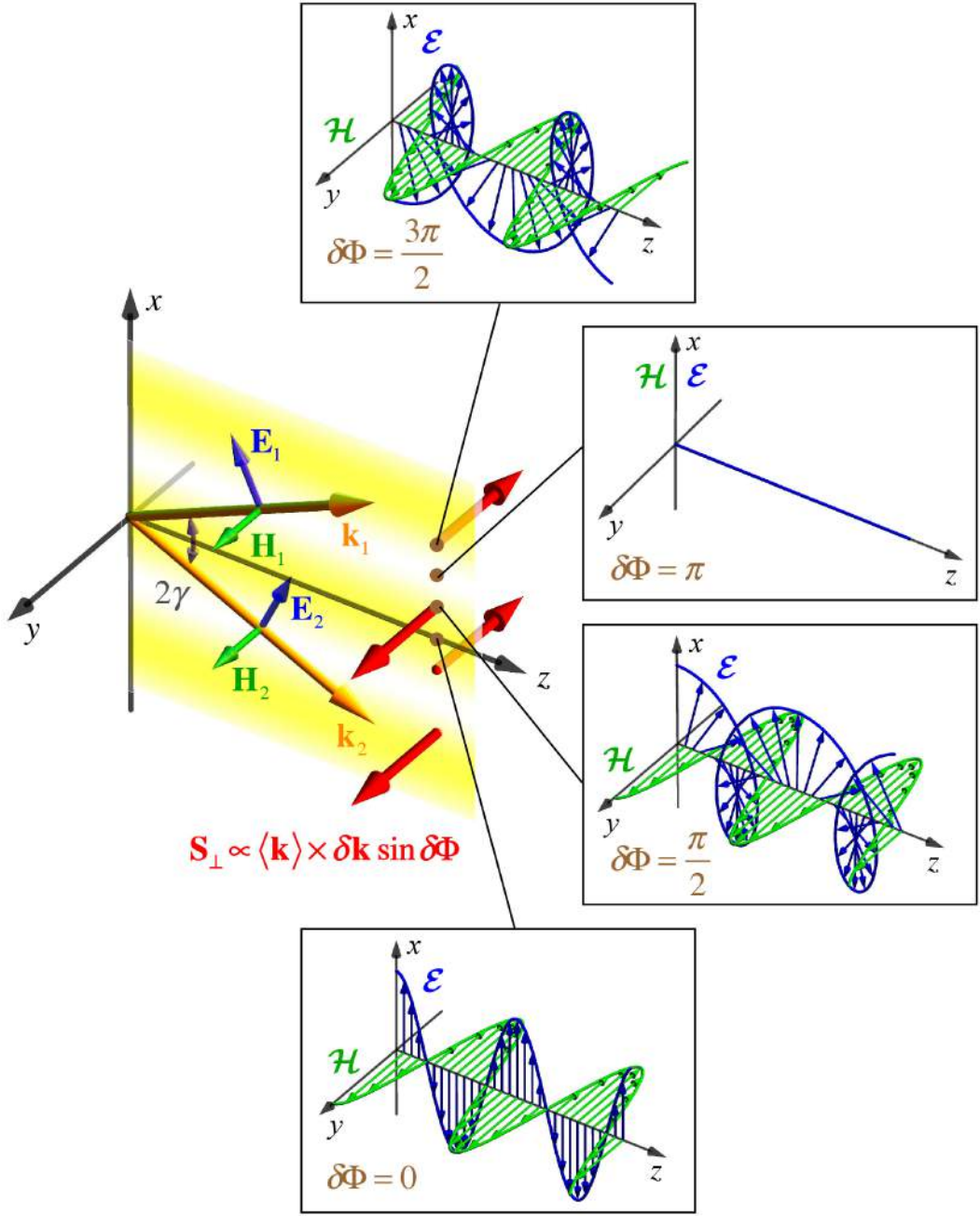

**Fig. 7.** Interference of two propagating plane waves (3.9)–(3.12) and the transverse (out-of-plane) spin AM density (3.14) or (3.16) [54], cf., Fig. 4. The simplest case of linear in-plane polarization $m=0$, $\tau=1$ is depicted. The inset panels display the $z$-evolutions of the instantaneous electric and magnetic fields $\mathcal{E}(\mathbf{r},0)$ and $\mathcal{H}(\mathbf{r},0)$ for different $x$-points of the interference pattern (marked by the phase difference $\delta\Phi$ between the waves). The cycloidal rotations of the electric field in the propagation $(x,z)$-plane generate the helicity-independent transverse spin AM density (3.14) and (3.16) $\mathbf{S}_\perp$. It has the properties listed in (VII) and varies sinusoidally across the interference pattern.

The resulting interference field is $\mathbf{E} = \mathbf{E}_1 + \mathbf{E}_2$ and $\mathbf{H} = \mathbf{H}_1 + \mathbf{H}_2$, and the interference picture is characterized by the relative $x$-dependent phase $\delta\Phi = \Phi_1 - \Phi_2 = 2k_x x$. For the sake of simplicity, we first consider equal polarizations of the two waves and the corresponding Stokes parameters, Eq. (2.16):

$$m_1 = m_2 \equiv m, \quad (\tau_1, \chi_1, \sigma_1) = (\tau_2, \chi_2, \sigma_2) \equiv (\tau, \chi, \sigma). \tag{3.11}$$

In this case the total wave electric field becomes

$$\mathbf{E} = \frac{2A_0}{\sqrt{1+|m_{1,2}|^2}} \left( \frac{k_z}{k} \cos\frac{\delta\Phi}{2} \bar{\mathbf{x}} + m\cos\frac{\delta\Phi}{2} \bar{\mathbf{y}} - i\frac{k_x}{k} \sin\frac{\delta\Phi}{2} \bar{\mathbf{z}} \right) \exp(ik_z z). \tag{3.12}$$

Importantly, this field has the *"imaginary"* longitudinal component $E_z \propto -i\frac{k_x}{k}\sin\frac{\delta\Phi}{2}$, which is quite similar to the longitudinal component in the evanescent wave (3.2), but here it oscillates and changes sign across the interference pattern.

Substituting Eqs. (3.9)–(3.12) into the general Eqs. (2.8)–(2.12) and (2.20), we obtain the energy, momentum, spin AM, and helicity densities in the two-wave interference field:

$$W = 2g|A_0|^2 \omega \left( 1 + \frac{k_z^2}{k^2} \cos\delta\Phi \right), \quad \mathbf{P} = \frac{W}{\omega} k_z \bar{\mathbf{z}}, \tag{3.13}$$

$$\mathbf{S} = 2g|A_0|^2 \left[ \sigma\frac{k_z}{k}(1+\cos\delta\Phi)\bar{\mathbf{z}} + \boxed{\frac{k_x k_z}{k^2}\sin\delta\Phi\,\bar{\mathbf{y}}} \right], \tag{3.14a}$$

$$\mathbf{S}^{e,m} = g|A_0|^2 \left[ \sigma\frac{k_z}{k}(1+\cos\delta\Phi)\bar{\mathbf{z}} + \boxed{(1\pm\tau)\frac{k_x k_z}{k^2}\sin\delta\Phi\,\bar{\mathbf{y}}} \;\boxed{\mp\chi\frac{k_x}{k}\sin\delta\Phi\,\bar{\mathbf{x}}} \right], \tag{3.14b}$$

$$K = \frac{W}{\omega}\sigma. \tag{3.15}$$

Here, akin to Eq. (3.6), we separated the electric and magnetic contributions (2.20) in the spin AM density (3.14), $\mathbf{S} = \mathbf{S}^e + \mathbf{S}^m$, because they show the electric-magnetic asymmetry of the transverse spin.

Equations (3.13)–(3.15) exhibit basic features similar to those of an evanescent wave, Eqs. (3.3), (3.4), and (3.6). Namely, the canonical momentum density $\mathbf{P}$ is proportional to the energy density $W$ and the mean wave vector $\langle\mathbf{k}\rangle = (\mathbf{k}_1 + \mathbf{k}_2)/2 = k_z\bar{\mathbf{z}}$ (Fig. 8b). The spin AM density (3.14) contains the usual longitudinal term $S_z$ proportional to the helicity parameter $\sigma$ and varying together with the energy density $W$ and the helicity density $K$, Eq. (3.15). In addition, there are also two transverse terms in Eqs. (3.14), which are orthogonal to the field momentum $\mathbf{P}$.

First, this is the *transverse (out-of-plane) helicity-independent spin AM density $S_y$* (shown in the red frame), which shares all the main features of its evanescent-wave counterpart in Eqs. (3.4)–(3.7) (see Figs. 7 and 8a). Namely, it is orthogonal to the momentum and the wave vectors $\mathbf{k}_{1,2}$, independent of the helicity and polarization, and is determined solely by the wave-vector (momentum) parameters. Second, akin to Eq. (3.6), the spin AM density (3.14) exhibits strong electric-magnetic asymmetry, and the *"dual-antisymmetric transverse (in-plane) spin AM"* densities $S_x^e = -S_x^m$ appear (shown in the green frame). These densities are directed along the wave inhomogeneity and are controlled by the second Stokes parameter $\chi$, exactly as described in Section 3.1.2 for evanescent waves.

The main difference between the transverse spin AM densities in the evanescent wave (3.6) and in the two-wave interference (3.14) is that the latter transverse spin oscillates and changes its sign across the interference pattern: $S_{x,y}^{e,m} \propto \sin\delta\Phi = \sin(2k_x x)$. Therefore, *the integral (i.e., the $\delta\Phi$-averaged) values of these transverse spins vanish*: $\langle \mathbf{S}_\perp^{e,m} \rangle = 0$. Nonetheless, locally they exist, and appear from the similar rotations of the field as in the evanescent wave. Figure 7 shows the instantaneous distributions of the real electric and magnetic fields in two-wave interference with the simplest linear TM polarization: $m = 0$, $\tau = 1$. One can see that the transverse (out-of plane) spin $S_y$ appears on the slopes of the interference picture due to the cycloidal rotation of the electric field in the propagation $(x,z)$-plane, see Eq. (3.12). Figure 8 shows the $(x,z)$-plane distributions of the electric-field polarization underlying the transverse spin density $S_y = S_y^e$, as well as the distribution of the canonical momentum $\mathbf{P}$. According to the general equations in Section 2.4, these quantities can be directly measured using local light-matter interactions: e.g., via the optical torque (2.23) and radiation-pressure force (2.22) on a probe particle.

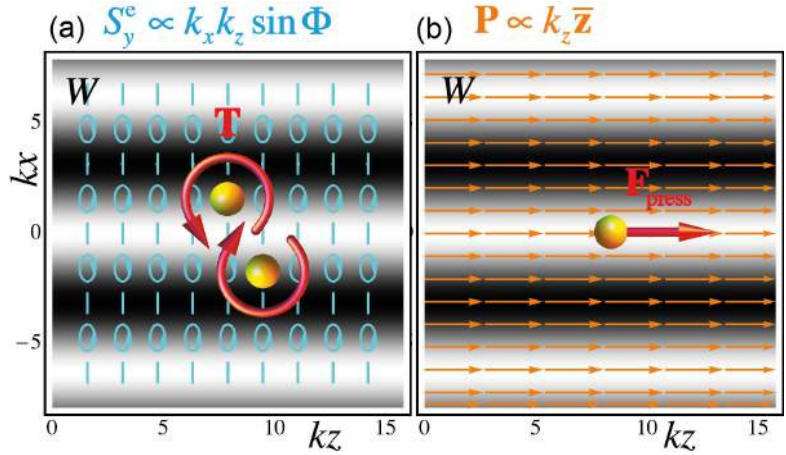

**Fig. 8.** The energy, momentum, and transverse spin AM densities (3.13) and (3.14) in the two-wave interference field (3.9)–(3.12) with $\gamma = 30°$ and linear $x$-polarization ($m = 0$, $\tau = 1$, see Fig. 7). The energy density $W(x)$ is shown as the grayscale background distribution. **(a)** The transverse spin AM density (3.14) $S_y^e(x)$ [normalized by $W(x)$] is shown in the form of a polarization-ellipse distribution in the propagation $(x,z)$-plane. It exerts the local transverse torque (2.23) $T_y$ on a probe particle [54]. **(b)** The longitudinal momentum density $\mathbf{P}$ is naturally proportional to the mean wave vector $\langle \mathbf{k} \rangle = k_z \bar{\mathbf{z}}$ and exerts the radiation-pressure force (2.22) $F_z$.

Using the mean wave vector $\langle \mathbf{k} \rangle$ and the wave-vector difference $\delta\mathbf{k} = \mathbf{k}_1 - \mathbf{k}_2 = 2k_x \bar{\mathbf{x}}$, we write the spin AM density (3.14) in the general vector form:

$$\mathbf{S} = 2g|A_0|^2 \left[ \sigma \frac{\langle \mathbf{k} \rangle}{k}(1 + \cos\delta\Phi) + \boxed{\frac{\langle \mathbf{k} \rangle \times \delta\mathbf{k}}{2k^2} \sin\delta\Phi} \right], \qquad (3.16a)$$

$$\mathbf{S}^{e,m} = g|A_0|^2 \left[ \sigma \frac{\langle \mathbf{k} \rangle}{k}(1+\cos\delta\Phi) + (1\pm\tau)\frac{\langle \mathbf{k} \rangle \times \delta\mathbf{k}}{2k^2}\sin\delta\Phi \mp \chi \frac{\delta\mathbf{k}}{2k}\sin\delta\Phi \right], \tag{3.16b}$$

One can note the remarkable similarity between Eq. (3.16) and Eq. (3.5), (3.7). This indicates that both types of the transverse spin mentioned above are rather generic phenomena. However, instead of the imaginary wave vector $\text{Im}\,\mathbf{k}$ in the evanescent-wave spin (3.5) and (3.7), the transverse terms in the two-wave spin (3.16) contain the real wave-vector difference $\delta\mathbf{k}$ and the phase difference $\delta\Phi$. Since the phase is a $\mathcal{P}$-even and $\mathcal{T}$-odd quantity, the product $\delta\Phi\delta\mathbf{k}$ has the same $\mathcal{P}$-odd and $\mathcal{T}$-even properties as the imaginary momentum $\text{Im}\,\mathbf{k}$ in the evanescent field. This ensures that all the terms in Eq. (3.16) have the proper AM symmetries.

Akin to the properties (V) and (VI), we now list the main features of the transverse spin AM densities in the interference field:

**Interfering waves AM:** Spin, Transverse (out-of-plane), Local.
Key parameters: $\langle \mathbf{k} \rangle, \delta\mathbf{k}, \delta\Phi$. (VII)

**Interfering waves AM:** Spin, Transverse (in-plane), Local, Anti-dual.
Key parameters: $\chi, \delta\mathbf{k}, \delta\Phi$. (VIII)

*3.2.2. Integral transverse (in-plane) helicity-dependent spin.* Alongside the local densities (3.13)–(3.16), we examine the *integral* dynamical properties of the two-wave interference field. Here, by integral we imply the natural averaging of the densities over the interference pattern, i.e.: $\langle ... \rangle \equiv \frac{1}{2\pi}\int_0^{2\pi} ... d(\delta\Phi)$. We now abandon the condition (3.11) and consider now the generic case of *different* polarizations $m_1$ and $m_2$ with their corresponding Stokes parameters (2.16). Performing straightforward calculations with the fields (3.10) and general equations (2.8)–(2.12), we derive (omitting the common $g|A_0|^2$ factor):

$$\langle W \rangle \propto 2\omega, \quad \langle \mathbf{P} \rangle \propto 2k_z\bar{\mathbf{z}} = \mathbf{k}_1 + \mathbf{k}_2, \tag{3.17}$$

$$\langle \mathbf{S} \rangle \propto \left[ \frac{k_x}{k}(\sigma_1-\sigma_2)\bar{\mathbf{x}} + \frac{k_z}{k}(\sigma_1+\sigma_2)\bar{\mathbf{z}} \right] = \sigma_1\frac{\mathbf{k}_1}{k} + \sigma_2\frac{\mathbf{k}_2}{k}, \tag{3.18}$$

$$\langle K \rangle \propto \sigma_1 + \sigma_2. \tag{3.19}$$

These equations reveal several interesting features. First, we emphasize that the electric and magnetic contributions to Eqs. (3.17) and (3.18) are equivalent: $\langle W^e \rangle = \langle W^m \rangle$, $\langle \mathbf{P}^e \rangle = \langle \mathbf{P}^m \rangle$, and $\langle \mathbf{S}^e \rangle = \langle \mathbf{S}^m \rangle$. In other words, the integral quantities become *dual-symmetric* in propagating fields [32]. Second, the transverse spin AM densities of Eqs. (3.14) and (3.16) disappear in the integral spin (3.18): $\langle \mathbf{S}_\perp \rangle = 0$ for $\sigma_1 = \sigma_2 = \sigma$. Finally, one can notice that the values (3.17)–(3.19) precisely correspond to the *sum of two single-photon energies $\omega$, momenta $\mathbf{k}_1$ and $\mathbf{k}_2$, spin AM $\mathbf{S}_1 = \sigma_1\frac{\mathbf{k}_1}{k}$ and $\mathbf{S}_2 = \sigma_2\frac{\mathbf{k}_2}{k}$, as well as helicities $\sigma_1$ and $\sigma_2$*. Thus, the $\delta\Phi$-averaging eliminates the fine interference features in the dynamical properties of the field and makes them *additive*.

Despite the simplicity of Eqs. (3.17)–(3.19), they offer *another type of transverse spin AM* in Eq. (3.18) (shown in the blue frame). Introducing the mean helicity $\langle\sigma\rangle = (\sigma_1 + \sigma_2)/2$ and the helicity difference $\delta\sigma = \sigma_1 - \sigma_2$, we represent Eq. (3.18) as

$$\langle\mathbf{S}\rangle \propto \boxed{\delta\sigma\frac{\delta\mathbf{k}}{2k}} + 2\langle\sigma\rangle\frac{\langle\mathbf{k}\rangle}{k}. \tag{3.20}$$

In the case of opposite helicities of the two interfering waves, $\sigma_1 = -\sigma_2 \equiv \sigma$, the mean helicity vanishes, $\langle K\rangle = \langle\sigma\rangle = 0$, and the integral spin AM (3.20) becomes *purely transverse* with respect to the momentum $\langle\mathbf{P}\rangle \propto \langle\mathbf{k}\rangle$, but *in-plane* with respect to the two wave vectors: $\langle\mathbf{S}\rangle \propto \sigma\,\delta\mathbf{k}$ (see Fig. 9). The idea of the transverse AM according to Eqs. (3.18) and (3.20) was first suggested by Banzer *et al.* in [49]. Figure 9 illustrates the appearance of the integral in-plane transverse spin AM (3.18) and (3.20) in the interference of circularly-polarized waves with $\sigma_1 = -\sigma_2 = 1$. One can see that this spin AM is generated by the cycloidal rotation of *both* the electric and magnetic fields in the $(y,z)$ plane.

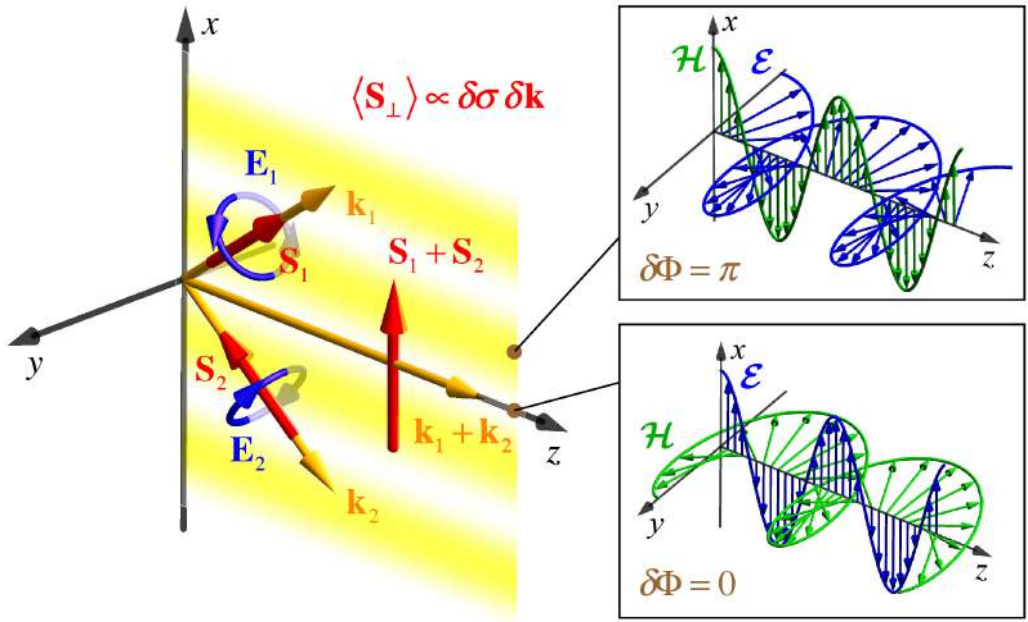

**Fig. 9.** Interference of two plane waves (3.9) and (3.10) with opposite circular polarizations ($m_1 = -m_2 = i$, $\sigma_1 = -\sigma_2 = 1$ here) and the transverse (in-plane) helicity-dependent spin AM (3.18) and (3.20) [49,54]. The integral ($\delta\Phi$-averaged) momentum $\langle\mathbf{P}\rangle$ and spin AM $\langle\mathbf{S}\rangle$ are obtained by simple summation of the single-wave momenta $\mathbf{k}_1$ and $\mathbf{k}_2$ and spins $\mathbf{S}_1 = \sigma_1\mathbf{k}_1/k$ and $\mathbf{S}_2 = \sigma_2\mathbf{k}_2/k$. The $z$-evolutions of the instantaneous electric and magnetic fields $\mathcal{E}(\mathbf{r},0)$ and $\mathcal{H}(\mathbf{r},0)$ in different $x$-points (marked by the $\delta\Phi$ values) exhibit the cycloidal rotations of both fields in the $(y,z)$-plane (cf., Fig. 7).

We emphasize that the *integral transverse (in-plane) spin AM* (3.18) and (3.20) is crucially determined by the *helicities* of the interfering waves. Unlike the previous transverse spins in Sections 3.1 and 3.2.1, it *cannot* appear in the interference of linearly-polarized waves with $\sigma_1 = \sigma_2 = 0$. Rather it represents a transversely-directed *perturbation* of the usual longitudinal spin AM (2.6) and (2.18), where $\sigma$ and $\mathbf{k}$ are substituted by their variations. Summarizing the properties of this third type of the transverse spin:

Here we have emphasized its non-vanishing integral character; but, of course, this spin also appears in local spin AM densities [54].

The equations (V) and (VII), (VI) and (VIII), as well as (IX), and the corresponding red-framed, green-framed, and blue-framed terms in the other equations describe *three distinct types of the transverse spin AM* in optical fields. These types differ in their nature, direction with respect to the wave vectors, and dependences on the polarization and wave-vector parameters. Their simple forms and appearance in very basic optical fields suggest that these three types of the transverse spin AM have a universal and robust nature. Therefore, it is natural to expect the presence of similar transverse spin densities in a variety of more complicated structured fields.

3.2.3. Focused Gaussian beam. The interference between two plane waves can serve as a toy planar model for *focused (non-paraxial) beams*, which consist of multiple plane waves propagating in different directions. We consider the simplest case of a focused polarized Gaussian beam. The transverse electric and magnetic fields of such beam, $\mathbf{E}_\perp$ and $\mathbf{H}_\perp$, can be taken from Eqs. (2.15) for paraxial beams, with $\ell = 0$ and the Gaussian envelope

$$A(\rho,z) = A_0 \frac{z_R}{q(z)} \exp\left(ik\frac{\rho^2}{2q(z)}\right). \tag{3.21}$$

Here $q(z) = z - iz_R$ is the complex beam parameter, $z_R = kw_0^2/2$ is the Rayleigh diffraction length, and $w_0$ is the beam waist [163]. In contrast to the paraxial approximation (2.15), we need to take into account non-zero *longitudinal* fields $E_z$ and $H_z$. In the first post-paraxial approximation (assuming $kz_R \gg 1$) these components can be determined from the transversality conditions $\nabla \cdot \mathbf{E} = \nabla \cdot \mathbf{H} = 0$ as

$$E_z \simeq ik^{-1}\nabla_\perp \cdot \mathbf{E}_\perp = -\frac{\rho}{q(z)}E_\rho, \quad H_z \simeq -\frac{\rho}{q(z)}H_\rho, \tag{3.22}$$

where $E_\rho$ and $H_\rho$ are the radial field components.

Combining the transverse field components (2.15) with the longitudinal fields (3.22), we obtain the full 3D field of the focused Gaussian beam:

$$\mathbf{E} = \frac{\overline{\mathbf{x}} + m\overline{\mathbf{y}} - \frac{x+my}{q(z)}\overline{\mathbf{z}}}{\sqrt{1+|m|^2}} A(\rho,z)e^{ikz}, \quad \mathbf{H} = \frac{\overline{\mathbf{y}} - m\overline{\mathbf{x}} - \frac{y-mx}{q(z)}\overline{\mathbf{z}}}{\sqrt{1+|m|^2}} A(\rho,z)e^{ikz}. \tag{3.23}$$

In the focal plane $z = 0$, $q = -iz_R$, and fields (3.23) acquire *"imaginary" longitudinal components*, which generate in-plane cycloidal rotations of the fields and the transverse spin AM density [55,150,151], Figs. 10 and 11.

Akin to the two-wave interference in Figs. 7 and 8, we consider the simplest $x$-linear polarization with $m = 0$, $\tau = 1$. The small longitudinal field components $E_z$ and $H_z$ make only second-order contributions to the energy, momentum, and helicity densities in the beam. Therefore, these are described by the paraxial Eqs. (2.17):

$$W \simeq g|A(\rho,z)|^2 \omega, \quad \mathbf{P} \simeq \frac{W}{\omega}k\overline{\mathbf{z}}, \quad K \simeq \frac{W}{\omega}\sigma = 0. \tag{3.24}$$

In contrast, the spin AM density (2.11) involves the first-order products between the longitudinal and transverse field components, which generate the *transverse spin AM densities* of both electric and magnetic origins:

$$\mathbf{S}^e = \frac{W}{\omega}\frac{x\,z_R}{z^2+z_R^2}\bar{\mathbf{y}}, \qquad \mathbf{S}^m = -\frac{W}{\omega}\frac{y\,z_R}{z^2+z_R^2}\bar{\mathbf{x}}. \tag{3.25}$$

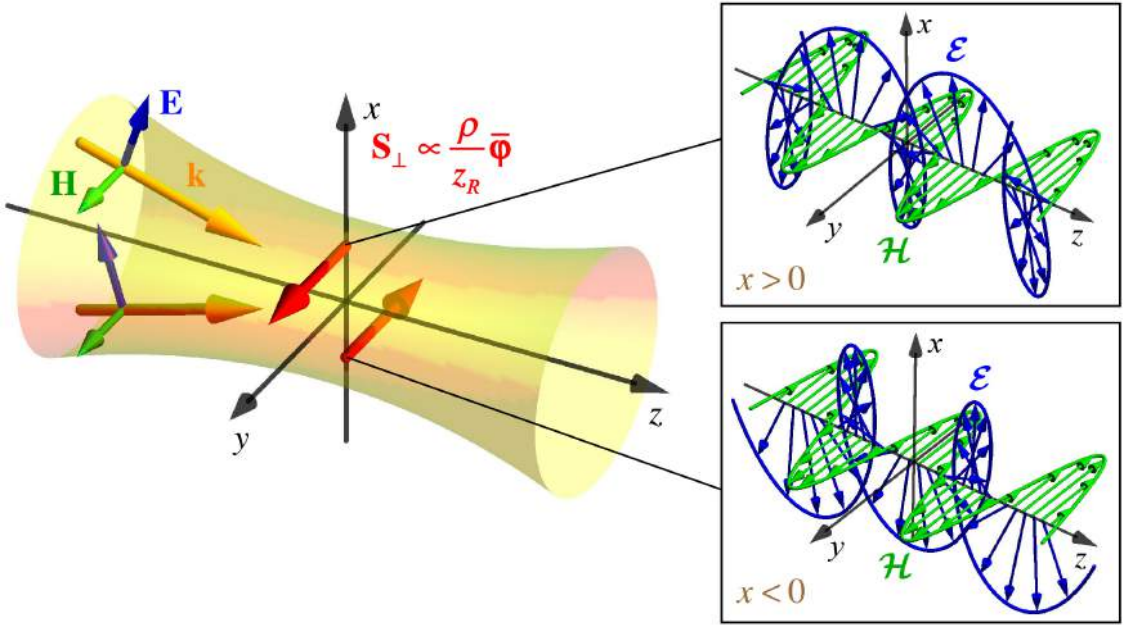

**Fig. 10.** Focused Gaussian beam (3.23) and the transverse (out-of-plane) spin AM density (3.25) and (3.26) near its focus [55,150,151]. The case of linear $x$-polarization ($m=0$, $\tau=1$) is shown. The transverse spin AM density $\mathbf{S}_\perp$ near the focus originates from the interference of plane waves forming the beam, cf., Fig. 7. The insets show the $z$-evolutions of the instantaneous electric and magnetic fields, $\mathcal{E}(\mathbf{r},0)$ and $\mathcal{H}(\mathbf{r},0)$, in the $x>0$ and $x<0$ halves of the beam near the beam focus. The opposite cycloidal rotations of the electric field generate the opposite-sign transverse spin AM density (3.25) $S_y^e$ in the upper and lower halves of the beam. The total (electric plus magnetic) transverse spin density (3.26) is independent of the beam polarization and is directed azimuthally: $\mathbf{S}_\perp \propto \bar{\boldsymbol{\varphi}}$.

These helicity-independent transverse spin AM densities are entirely similar to those considered in the two-wave interference [the red-framed terms in Eqs. (3.14) and (3.16)]. For instance, considering the $y=0$ cross-section of the beam (Figs. 10), we see that the transverse (out-of-plane) spin AM density $S_y = S_y^e$ is of purely electric origin and appears due to the interference of plane waves forming the beam in the $(x,z)$ plane. This spin density is maximal in the focal plane $z=0$ and has opposite signs in the $x>0$ and $x<0$ halves of the beam, exactly like the opposite spin densities on the two slopes of an interference fringe in Figs. 7 and 8. The transverse spin AM (3.25) is characterized by properties (VII) of the transverse (out-of-plane) spin in interfering waves. Indeed, the inverse Rayleigh-length parameter $w_0/z_R$ is proportional to the variations in the transverse wave-vector components in the beam spectrum, $\delta k_{x,y}$, whereas the dimensionless $x/w_0$ and $y/w_0$ distances from the beam axis underpin the transverse phases $\delta\Phi_{x,y}$ in the wave interference (constructive interference on the beam axis and a destructive one at infinity). Figure 11 shows the $(x,z)$ distributions of the in-plane polarization ellipses (with their ellipticity proportional to the transverse spin AM density $S_y^e$) and momentum density $\mathbf{P}$ in the focused Gaussian beam (3.23). The presence of the in-plane elliptical polarizations near the focus of such beams was recently emphasized by Yang and Cohen [150] and measured in [55,151] (see Subsection 3.3.4).

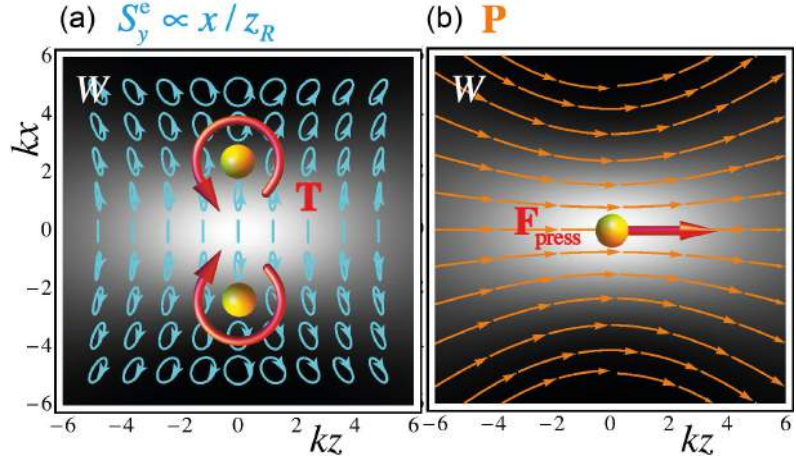

**Fig. 11.** The energy, momentum, and transverse spin AM densities in the focused Gaussian beam (3.23) with $kz_R = 5$ and linear $x$-polarization ($m = 0$, see Fig. 10). The energy density $W(x)$ is shown as the grayscale background distribution. **(a)** The transverse spin AM density (3.25) $S_y^e(x)$ [normalized by $W(x)$] is shown in the form of the electric-field polarization distribution in the $(x,z)$-plane [55,150,151]. It exerts the local transverse torque (2.23) $T_y$ on a probe particle. **(b)** The longitudinal momentum density (2.9) **P** has the main longitudinal component proportional to $k$, a small radial component due to the diffraction, and it exerts the radiation-pressure force (2.22) $F_z$.

Note that the total (electric plus magnetic) transverse spin (3.25) is directed azimuthally, i.e., orthogonally to the radial wave-vector distribution. This polarization-independent (for uniformly-polarized beams) spin AM can be written as

$$\mathbf{S}_\perp = \frac{W}{\omega} \frac{\rho z_R}{z^2 + z_R^2} \bar{\boldsymbol{\varphi}} \quad . \tag{3.26}$$

Thus, a focused beam exhibits a *transverse-spin vortex* in its focal plane (Fig. 10). Its direction is determined solely by the radial inhomogeneity and propagation direction of the beam (cf., the edge transverse spin in Section 3.3.5 and [56]). Obviously, the transverse spin (3.26) does not contribute to the integral spin AM: $\langle \mathbf{S}_\perp \rangle = 0$.

### *3.3. Measurements and applications*

We are now in a position to describe the main experimental measurements involving the transverse spin AM. The transverse spin densities were recently detected in both evanescent and propagating fields. Moreover, due to its unusual properties (V), the transverse (out-of-plane) spin AM of evanescent waves, Eq. (3.5), has been employed for the robust spin-dependent transport of light.

<u>3.3.1. Reconstructions of 3D fields.</u> First of all, the elliptical in-plane polarization in evanescent waves has been known for a long time in the literature; it can be found in textbooks and reviews on evanescent waves and plasmonics (see, e.g., [155,157,160]). In the past decade, rapid progress in nano-optics stimulated the development of several methods allowing the probing and reconstruction of full 3D polarization distributions in structured optical fields [161,164–167]. The full 3D polarization at a given point of a structured optical field can be reconstructed using local subwavelength probes: e.g., a small nanoparticle scatterer [161,167] or a near-field tip [164–166]. In this manner, the elliptical in-plane polarization in evanescent waves was measured in [161], Fig. 12a, as well as similar polarizations in non-paraxial (focused) beams in [166], Fig. 12b. Note that rotating in-plane magnetic field in Fig. 12b corresponds to the transverse magnetic spin AM $S_x^m \propto +y$ or $\mathbf{S}_\perp \propto -\rho\bar{\varphi}$, i.e., of opposite sign as compared with Eqs. (3.25) and (3.26). This is because of the non-uniform azimuthal polarization in the beam, which results in the addition $\pi$ phase in the interference of pairs of plane waves in the beam (cf., Figs. 7 and 10). Obviously, such full reconstructions of the electric and magnetic wave fields allow retrieving all the properties of these fields (including momentum, spin, etc.), so that probing the 3D field can be used as an *indirect* measurement of the dynamical properties of the field.

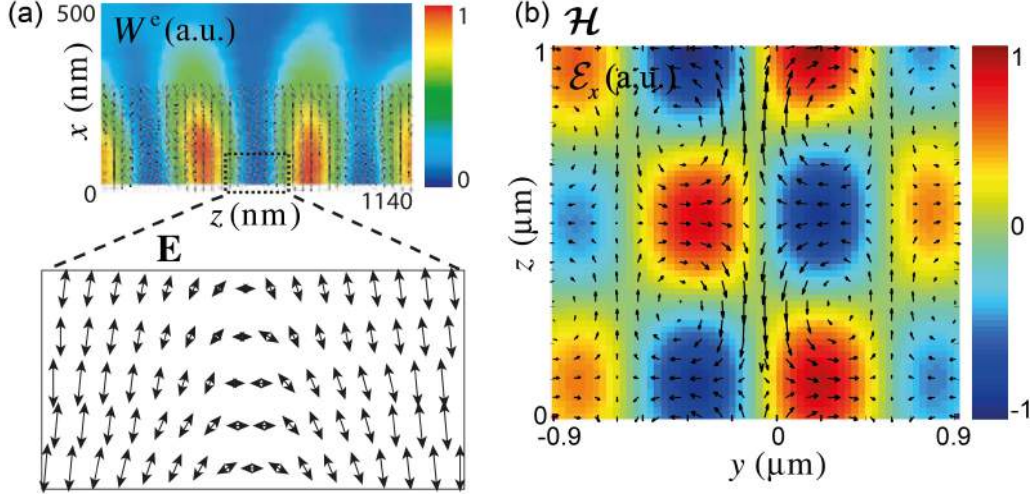

**Fig. 12. (a)** Reconstruction of the $(x,z)$-polarization of the electric field $\mathbf{E}(\mathbf{r})$ [shown by arrows] of a *standing* evanescent wave via probe-particle scattering [161]. The background color-scale plot depicts the electric energy density $W^e(\mathbf{r})$. Rotation of the linear polarization with $z$ is a signature of elliptical polarizations in the counter-propagating evanescent waves forming the standing wave, cf. Figs 4 and 5a. **(b)** Reconstruction, via a near-field probe, of the $(y,z)$-components of the instantaneous magnetic wave field $\mathcal{H}(\mathbf{r},0)$ [shown by arrows] in a $z$-propagating azimuthally-polarized Bessel beam [166]. The background color-scale plot shows the $x$-component of the instantaneous electric field $\mathcal{E}(\mathbf{r},0)$. Note the opposite rotations of the magnetic field in the $y > 0$ and $y < 0$ halves of the beam, which indicate the transverse magnetic spin AM density $S_x^m \propto y$ (a radial polarization would result in the analogous electric spin AM density, cf. Figs. 10, 11a and Fig. 14 below).

3.3.2. Optomechanical calculations and measurements. A *direct* detection of the dynamical field characteristics involves *optomechanical* methods. The straightforward mechanical detection of the spin AM density employs small probe particles and an optical torque acting on these (see Fig. 3) [2,16,17,50,54,134]. Interestingly, already in 1998 Chang and Lee [162] calculated optical torques on a spherical particle in an evanescent field and found an unusual transverse torque $T_y$, which was independent of the wave helicity. However, they interpreted this torque as coming from the vertical $x$-gradient of the longitudinal $z$-directed optical pressure. In fact, this was the torque (2.23) originating from the transverse spin AM density $S_y^e$, Eqs. (3.4)–(3.7). One can show that for small subwavelength particles of radius $a$, $ka \ll 1$, the gradient-pressure torque (suggested by Chang and Lee) is much smaller than the transverse-spin torque. Complete calculations of all the three components of the optical torque $\mathbf{T}$ acting on a spherical particle in an evanescent wave (3.2) were made in [50], see Figs. 13a,b. In perfect agreement with Eqs. (2.23) and (3.6) or (3.7), these calculations clearly show: (i) the longitudinal $\sigma$-dependent torque $T_z$ due to the usual spin AM $S_z^e$; (ii) the $\tau$-dependent (but helicity-independent) transverse torque $T_y$, originating from the electric transverse (out-of plane) spin $S_y^e$; and (iii) the vertical $\chi$-dependent torque $T_x$ produced by the transverse (in-plane) electric spin AM density $S_x^e$ (despite the vanishing total vertical spin (3.4) $S_x = 0$). Importantly, the dependences of these torques on the polarization Stokes parameters hold true *exactly* even for larger Mie particles with $ka > 1$, i.e., beyond the range of validity of the dipole-approximation Eq. (2.23). This confirms that the spin AM explanation of the mechanical action of light is robust and fundamental.

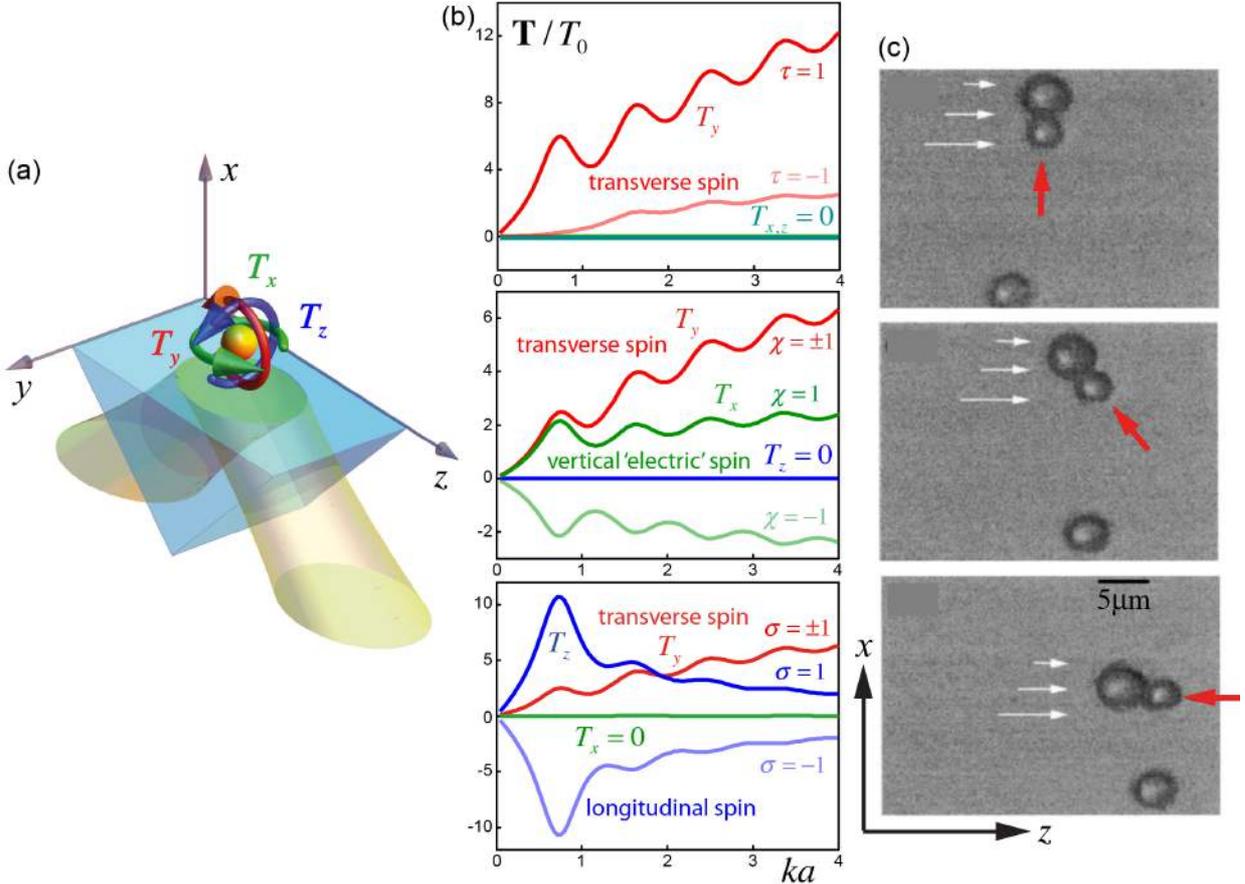

**Fig. 13.** Optical torques for a particle immersed in an evanescent field. **(a)** Schematics of an experiment with the evanescent field (3.2) generated by the total internal reflection and three components of the torque **T** indicated by rotational directions. **(b)** Numerical calculations [50] of the three components of the torque **T** on a gold spherical particle of radius $a$ for six basic polarizations of the evanescent wave, which are described by the Stokes parameters $(\tau, \chi, \sigma) = \pm 1$. The $\sigma$-dependent longitudinal torque $T_z$, $\tau$-dependent transverse torque $T_y$, and the $\chi$-dependent vertical torque $T_x$ perfectly correspond to the three terms in the electric spin AM density $\mathbf{S}^e$, Eqs. (3.6) and (3.7), even beyond the dipole-approximation ($ka \ll 1$) of Eq. (2.23). **(c)**. Experimental observation of the transverse (out-of-plane) torque $T_y$ via rotation of a double Mie particle [168]. Due to the complex macroscopic shape of the particle, the torque here can also be exerted by the vertical $x$-gradient of the longitudinal ($z$) radiation-pressure force (2.22) (indicated by white arrows), as it was interpreted in [162,168]. In the limit of small subwavelength particles, only the torque (2.23) from the spin AM density survives.

In 1999, Song *et al.* [168] reported the first experimental observation of the transverse in-plane rotation of probe Mie particles in evanescent waves, Fig. 13c, again interpreted as the vertical gradient of the longitudinal radiation pressure. In that experiment such mechanism was indeed possible (in addition to the transverse spin), because large clusters of Mie particles with $ka \gg 1$ were used. Therefore, although the experiment [168] can be considered as the first optomechanical confirmation of the transverse (out-of-plane) spin AM in evanescent waves (without recognizing it), additional experiments with nanoparticles would be desirable. Finally, optomechanical attempt to detect the transverse (in-plane) $\chi$-dependent electric spin AM (3.6) and (3.7), as well as the transverse spin AM densities in propagating fields (Section 3.2.1) are still to be done.

3.3.3. *Probing optical spin using atoms in a magnetic field.* The light-atom interaction can be used as a quantum-mechanical analogue of classical particles probing light. In this manner, transitions between different atomic angular-momentum states correspond to the torque on a classical particle [29,130]. These internal angular-momentum atomic states are characterized by the $m$ quantum number (electron vortex number) and are defined with respect to a certain quantization axis. To fix the quantization axis and distinguish different $m$-states, usually an external static magnetic field $\mathbf{H}_0$ is introduced, which generates a fine Zeeman splitting of the atomic $m$-sublevels [152–154]. Thus, atomic transitions involving different $m$-levels with respect to the corresponding axis can probe the local polarization and spin state of an optical field.

In this manner, the first attempt to perform the atomic Zeeman spectroscopy of the evanescent-wave polarization was made in [153], where transitions to $m = \pm 1$ and $m = 0$ levels for different directions of an external magnetic field were measured. These measurements clearly indicated the longitudinal $z$-component in the evanescent-field polarization for $\mathbf{H}_0 = H_0 \overline{\mathbf{z}}$ (see Fig. 8c in [153]), but could not properly show the ellipticity of the polarization in the $(x,z)$-plane for $\mathbf{H}_0 = H_0 \overline{\mathbf{y}}$ (see Fig. 8a in [153]).

Recently, a series of atomic measurements by Mitsch *et al.* [52,57,154] perfectly confirmed the presence of the elliptical polarization and transverse spin in evanescent waves. Generating evanescent waves with opposite directions of propagations $\mathrm{Re}\,\mathbf{k} \propto \pm \overline{\mathbf{z}}$ or opposite decay directions $\mathrm{Im}\,\mathbf{k} \propto \pm \overline{\mathbf{x}}$, experiments [52,57,154] indicated the presence of the corresponding transverse spin AM (3.5) $\mathbf{S}_\perp \propto \pm \overline{\mathbf{y}}$. In particular, two sorts of experiments were realized in [154]. In both experiments, a homogeneous external magnetic field $\mathbf{H}_0 = H_0 \overline{\mathbf{y}}$ was applied. First, in this setting, a resonant evanescent optical field drives $\Delta m = +1$ ($\Delta m = -1$) transitions when its transverse spin is parallel (antiparallel) to the external magnetic field. This was confirmed by optically pumping cold atoms to the $m = +4$ and $m = -4$ Zeeman states for the two signs of the transverse spin $S^\mathrm{e}_y$. Second, using an off-resonant interaction, where the spin AM of light acts as a fictitious magnetic field $\mathbf{H}_0^\mathrm{fict} \propto \mathbf{S}^\mathrm{e}$, the Zeeman splitting of levels proportional to $\left(\mathbf{H}_0 + \mathbf{H}_0^\mathrm{fict}\right)$ was detected, which clearly indicated the transverse spin $S^\mathrm{e}_y$ in evanescent waves.

3.3.4. *Magneto-optical and particle-scattering measurements.* Two remarkable measurements of the transverse spin AM density in focused Gaussian beams were recently realized in [55,151]. These works used completely different methods and confirmed the same results of Figs. 10, 11 and Eqs. (3.25), (3.26).

First, as we indicated in Section 2.4 and Eq. (2.27), the spin AM density $\mathbf{S}^\mathrm{e}$ is naturally coupled to an external magnetic field $\mathbf{H}_0$ in magnetic circular dichroism (MCD) [150]. Experiment by Mathevet and Rikken [151] used the $x$- and $y$-polarized focused Gaussian beams, i.e., $m = 0$, $\tau = 1$ and $m = \infty$, $\tau = -1$, respectively. The external magnetic field $\mathbf{H}_0 = H_0 \overline{\mathbf{y}}$ and MCD crystal probe were used to measure the differential MCD signal between the upper $x > 0$ and lower $x < 0$ halves of the beam (Fig. 14a,b). As a result, a clear MCD response was detected for the $x$-polarized but not the $y$-polarized beam. Thus, these measurements registered the presence of a purely *electric* transverse spin AM density (3.25) $S_y = S_y^e$ (with opposite signs in the $x > 0$ and $x < 0$ halves of the beam) for the $x$-polarization, while these were insensitive to the transverse magnetic spin AM density $S_y = S_y^m$ which takes place for the $y$-polarization. This is an important evidence of the dual (electric-magnetic) asymmetry of the transverse helicity-independent spin AM.

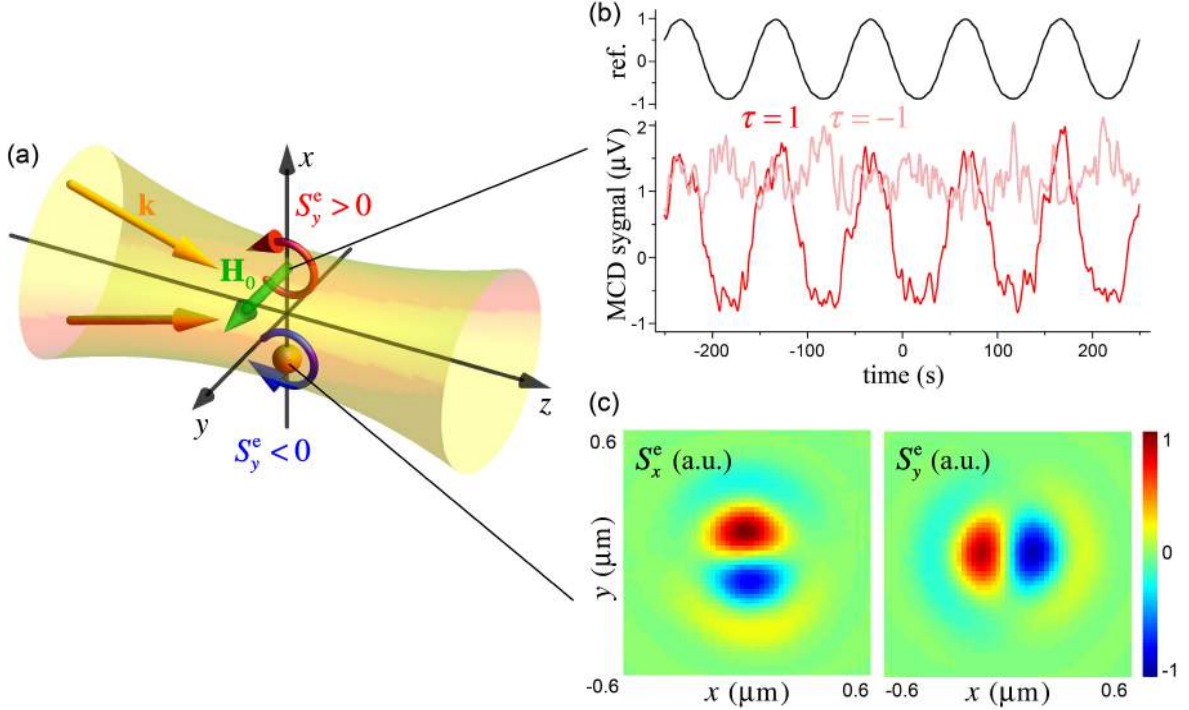

**Fig. 14.** Recent measurements of the transverse spin AM density (3.25) and (3.26) in a focused linearly-polarized Gaussian beam (cf. Figs. 10 and 11). **(a)** Schematics of the experiments [151] and [55], using magnetic circular dichroism (MCD) with the transverse static magnetic field $\mathbf{H}_0 = H_0 \overline{\mathbf{y}}$ and probe-particle scattering, respectively. **(b)** Experimentally measured MCD signal in the $x > 0$ half-beam (vs. the reference MCD signal shown in black) for the $x$- and $y$-polarizations ($\tau = \pm 1$) [151]. The presence of the $\tau = 1$ signal and absence of the $\tau = -1$ signal correspond to the presence of the electric spin AM density (3.25) $S_y^e$ in the first case and the magnetic spin AM density $S_y^m$ in the second case. The latter does not induce the MCD response, see Eq. (2.27). **(c).** Transverse electric spin AM distribution in the focal plane of a radially-polarized Gaussian beam, which was measured in [55] using nontrivial scattering-particle probing. In this case, the electric spin AM density has an azimuthal distribution (3.26): $\mathbf{S}_\perp = \mathbf{S}_\perp^e \propto -\overline{\boldsymbol{\varphi}}$ (cf., the azimuthal distribution of the magnetic spin AM density $\mathbf{S}_\perp = \mathbf{S}_\perp^m \propto -\overline{\boldsymbol{\varphi}}$ in the azimuthally-polarized beam in Fig. 12b).

Second, the experiment by Neugebauer *et al.* [55] extended their previous works [167,169] developing field scattering by a probe nanoparticle for detection of nontrivial properties of an optical field. Namely, they showed that asymmetries in the intensities of the particle-scattered *near*-fields are directly proportional to the transverse electric spin AM densities in the field. Employing this observation and scanning the focal plane of the beam with a probe nanoparticle, the experiment [55] provided a direct mapping of the transverse electric spin AM densities in linearly $x$-polarized and radially-polarized focused Gaussian beams, Fig. 14c. The results revealed the transverse $y$-directed spin AM density $S_y^e \propto +x$, Eq. (3.25), for the $x$-polarized beam, as well as the transverse azimuthal spin $\mathbf{S}_\perp = \mathbf{S}_\perp^e \propto -\rho \overline{\boldsymbol{\varphi}}$ in the radially-polarized beam.

Note that the azimuthal spin AM density has a purely electric (magnetic) origin for radially- (azimuthally-) polarized beams (cf. Fig. 14c and 12b). In addition, as we remarked in Section 3.3.1, the transverse spin in such non-uniformly polarized beams has the sign opposite to the uniformly-polarized case (3.26).

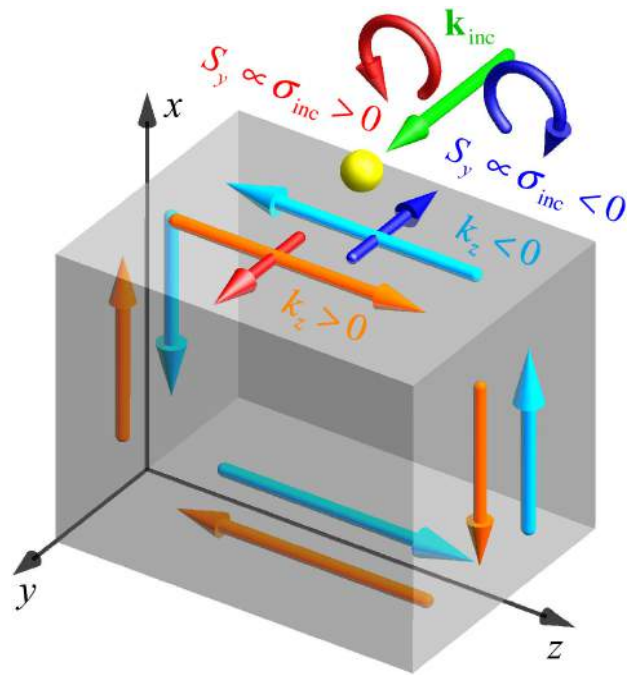

**Fig. 15.** The quantum spin Hall effect of light manifests itself as a strong transverse spin-momentum coupling (locking) in evanescent surface waves [56], see Eqs. (3.5) and (V). Any interface supporting surface waves with evanescent free-space tails (3.2) has counterpropagating modes [in orange and cyan] with opposite transverse spins [in red and blue]. This is observed experimentally via the excitation of unidirectional surface-evanescent modes with $k_z > 0$ or $k_z < 0$, depending on the helicity $\sigma_{inc} = 1$ or $\sigma_{inc} = -1$ and the corresponding $y$-directed longitudinal spin AM (2.17) of the transversely-incident light [in green] (see [52,53,74–83] and Fig. 16). The incident propagating light is coupled to the surface-evanescent modes via a scatterer [in yellow]: e.g., a nanoparticle or an atom.

3.3.5. Spin-directional coupling and quantum spin Hall effect of light. Perhaps the most remarkable application of the transverse spin AM is its ability to provide *spin-controlled unidirectional propagation of light* [52,53,74–83]. This is based on the property (V), which strongly couples the transverse spin AM of an evanescent wave, Eq. (3.5) and Fig. 4, with its direction of propagation. Indeed, evanescent waves propagating in opposite directions along the same interface (i.e., having opposite Re**k** but the same Im**k**) have opposite transverse spins (3.5), as shown in Fig. 15. In a similar manner, evanescent waves propagating in the same direction on opposite sides of the same sample (i.e., having the same Re**k** but opposite Im**k**) also have opposite transverse spins. Notably, these features are independent of the nature of the interface, and are valid for *any interfaces supporting evanescent waves* (3.2). These could be metallic surfaces with plasmon-polariton modes [53,74,75], optical nano-fibers [52,57,77], or photonic-crystal waveguides [78,80,81]. In all cases, 2D or 3D samples supporting edge or surface modes with evanescent tails (3.2) have *counter-propagating modes with opposite transverse spins* (3.5), Fig. 15. This is a fundamental property of free-space Maxwell equations that the transverse spin is strongly coupled (locked) with the direction of propagation of light. Recently it was shown that this property can be associated with the intrinsic *quantum spin Hall effect of light*, which originates from the spin-orbit coupling and topological properties of photons [56]. Analogous topological phenomena for electrons in solids recently gave rise to a new class of materials: *topological insulators* [170,171].

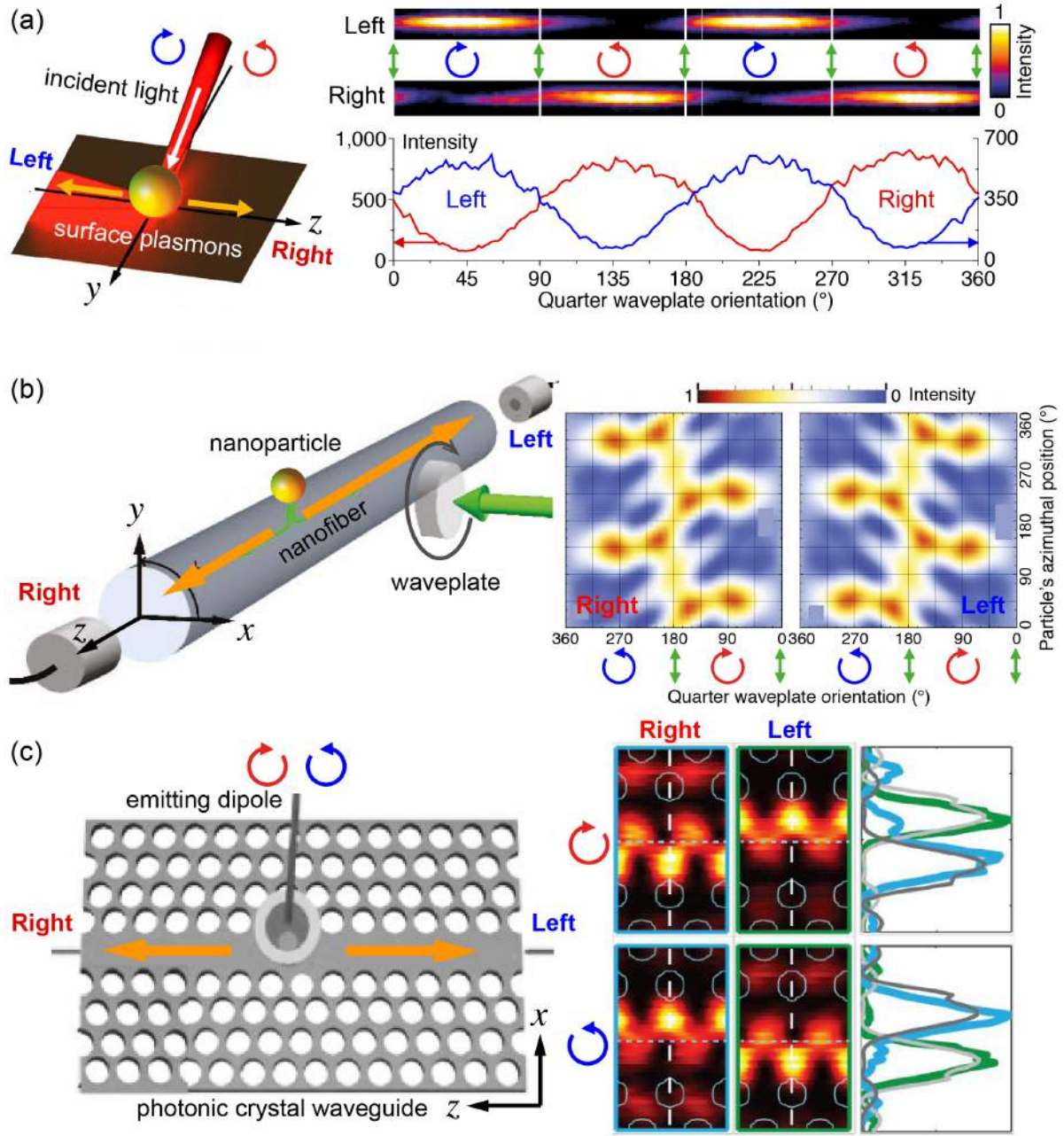

**Fig. 16.** Examples of recent experiments [53,77,78] demonstrating spin-controlled unidirectional excitation of modes with evanescent tails. Despite very different systems [surface plasmon-polaritons in **(a)**, nanofiber in **(b)**, and photonic-crystal waveguide **(c)**], all these experiments show the robust quantum spin Hall effect properties of light with the transverse spin-momentum locking in evanescent waves, see [56] and Fig. 15.

Transverse spin-momentum locking is attracting rapidly-growing attention, and a number of prominent experiments has been reported by different groups [52,53,57,74–80]. Figure 16 shows examples from several experiments using quite different setups and interfaces, but all based on the same transverse spin-direction coupling. In all these cases external $y$-propagating light ( $\mathbf{k}_{inc} = k\overline{\mathbf{y}}$ ) with usual longitudinal spin AM (2.17) from the circular polarization ( $\sigma_{inc} = \pm 1$ ) was coupled to surface-evanescent waves propagating along the $z$-axis via some scatterer (nanoparticle, atom, etc.). Since the spin AM of the incident wave has to *match* the $y$- directed transverse spin AM in evanescent waves, opposite helicities ($\sigma_{inc} = 1$ and $\sigma_{inc} = -1$) of the incident light generated unidirectional evanescent waves propagating in opposite $z$-directions ($k_z > 0$ and $k_z < 0$), respectively. Thus, such a strong and robust spin-direction locking offers perfect "chiral unidirectional interfaces", potentially important for many applications, such as quantum information [77,78], topological photonics [56,172], and chiral spin networks [83]. Note that the spin-direction coupling is *reversible*, i.e., the transverse emission of propagating light from oppositely-propagating evanescent waves has opposite circular polarizations (spins) [53]. Furthermore, combining transverse spin-direction locking with the spin-dependent magnetooptical scattering or absorption [52,154] (Section 3.3.3), results in an efficient "optical diode", i.e., non-reciprocal transmission of light [57].

To conclude this Section 3, various examples of the transverse spin AM densities in basic optical fields reveal themselves in a variety of experiments involving light-matter interactions. Remarkably, in many cases, the presence and key role of the transverse spin were not properly realized. Now, having the theoretical considerations and classification provided in this review, as well as an analysis of the experiments, one can properly appreciate the important role of the transverse spin AM as one of the inherent dynamical properties of light.

## 4. Transverse orbital angular momenta

In Section 3 we only considered the spin AM of light, which is determined by the intrinsic (polarization) degrees of freedom. We now consider the transverse orbital AM, determined by the spatial (phase) degrees of freedom.

### *4.1. Transverse extrinsic orbital AM*

4.1.1. General features. In Section 2 we described the main properties of the longitudinal orbital AM of light (IV), which is produced in paraxial vortex beams and is widely used in modern optics [3–14]. According to Eqs. (2.10), (2.17) and (2.18), such orbital AM is *extrinsic locally*, but becomes *intrinsic integrally*. (Recall that we use "intrinsic" and "extrinsic" to distinguish the coordinate origin-independent and origin-dependent quantities, respectively.) Therefore, we regard such integral longitudinal vortex-dependent AM as *intrinsic orbital AM*: $\langle \mathbf{L} \rangle = \langle \mathbf{L}^{int} \rangle \propto \ell \langle \mathbf{k} \rangle / k$.

The $z$-propagating vortex beams (2.15) represent eigenmodes of the quantum AM operator $\hat{L}_z = -i \frac{\partial}{\partial \varphi}$. Thus, they reveal the *wave* (intrinsic) features of the orbital AM. But what about the *particle* (extrinsic) aspects of the orbital AM, Eq. (2.1)? The paraxial-beam analysis in Section 2.3 was done based on an assumption that the $z$-axis coincides with the beam axis, i.e., the beam passes through the coordinate origin (Fig. 2). To unveil the particle-like AM of light, we have to consider the same paraxial beam (2.15) but now *shifted* away from the origin. Without loss of generality, we assume that the beam center is displaced by the vector $\mathbf{r}_0 = \Delta \bar{\mathbf{y}}$ along the $y$-axis, and the beam still propagates parallel to the $z$-axis, Fig. 17.

The beam field is given by Eq. (2.15) with the transformation (2.2):

$$\mathbf{r} \to \mathbf{r} + \mathbf{r}_0. \quad (4.1)$$

This transformation trivially translates the energy (2.8), momentum (2.9), spin AM (2.11), and helicity (2.12) distributions. Only the orbital AM density (2.10) explicitly involves the radius vector $\mathbf{r}$, and, thus, transforms as in the point-particle Eq. (2.3):

$$\mathbf{L} \to \mathbf{L} + \mathbf{r}_0 \times \mathbf{P}. \quad (4.2)$$

From this equation we immediately obtain the transformation of the *integral* orbital AM induced by the beam shift (4.1):

$$\langle \mathbf{L} \rangle \rightarrow \langle \mathbf{L} \rangle + \boxed{\mathbf{r}_0 \times \langle \mathbf{P} \rangle} . \tag{4.3}$$

Importantly, the second term in Eq. (4.3) (shown in the magenta frame) is *transverse* with respect to the mean momentum $\langle \mathbf{P} \rangle$, and also *extrinsic* because it explicitly involves position of the beam with respect to the coordinate origin. For the paraxial vortex beam (2.15), transformations (4.2) and (4.3) yield [cf. Eq. (2.17) and (2.18)]:

$$\mathbf{L} \simeq \frac{W}{\omega}(-\rho k \overline{\varphi} + \ell \overline{\mathbf{z}}) + \frac{W}{\omega} k \Delta \overline{\mathbf{x}}, \quad \langle \mathbf{L} \rangle \propto \ell \overline{\mathbf{z}} + \boxed{k \Delta \overline{\mathbf{x}}} \equiv \langle \mathbf{L}^{\text{int}} \rangle + \langle \mathbf{L}^{\text{ext}} \rangle. \tag{4.4}$$

Here the $\ell$-dependent terms describe the longitudinal intrinsic vortex-dependent orbital AM $\langle \mathbf{L}^{\text{int}} \rangle$, which remains unchanged as in Eqs. (2.17) and (2.18). At the same time, the $\Delta$-dependent terms in Eqs. (4.4) originate from the cross-product (4.3) between the transverse coordinate of the beam center, $\mathbf{r}_0$, and the longitudinal momentum of the beam, $\langle \mathbf{P} \rangle \propto k \overline{\mathbf{z}}$, see Fig. 17. This is the *extrinsic orbital AM* of light $\langle \mathbf{L}^{\text{ext}} \rangle = \mathbf{r}_0 \times \langle \mathbf{P} \rangle$, which is always *transverse* by its definition (4.3). This AM is independent of the vortex or polarization and is determined by the most basic *particle* properties of a light beam: its position and direction of propagation, i.e., *trajectory*.

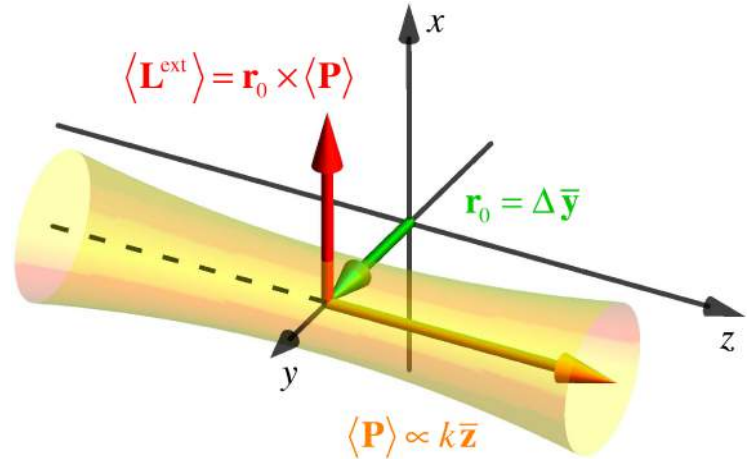

**Fig. 17.** The extrinsic orbital AM of light, Eqs. (4.1)–(4.5). The paraxial optical beam is shifted away from the coordinate origin by the $\mathbf{r}_0$ vector. It still carries the same momentum and intrinsic longitudinal AM as in Fig. 2, but also acquires the extrinsic orbital contribution $\langle \mathbf{L}^{\text{ext}} \rangle = \mathbf{r}_0 \times \langle \mathbf{P} \rangle$, which is transverse with respect to the mean momentum $\langle \mathbf{P} \rangle$. The extrinsic orbital AM takes into account the mechanical particle-like properties of light: its coordinate and momentum, i.e., its *trajectory*.

Since the orbital AM density (2.10) is extrinsic by its definition, the separation between the intrinsic and extrinsic orbital AM appears only in *integral* values. For the generic localized optical field, one can calculate the mean (expectation) values of the coordinate (weighted with the energy density (2.8) or another relevant optical density), $\langle \mathbf{r} \rangle$, momentum, $\langle \mathbf{P} \rangle \propto \langle \mathbf{k} \rangle$, and orbital AM as $\langle \mathbf{L} \rangle = \langle \mathbf{r} \times \mathbf{P} \rangle$. Then, the intrinsic and extrinsic parts of the orbital AM can be separated as [31]

$$\boxed{\langle \mathbf{L}^{\text{ext}} \rangle = \langle \mathbf{r} \rangle \times \langle \mathbf{P} \rangle}, \quad \langle \mathbf{L}^{\text{int}} \rangle = \langle \mathbf{L} \rangle - \langle \mathbf{L}^{\text{ext}} \rangle. \tag{4.5}$$

This definition coincides with Eq. (4.4) in the case of paraxial beams. Since both the local coordinates $\mathbf{r}$ and the expectation values $\langle \mathbf{r} \rangle$ are equally transformed upon translations (4.1), equations (4.5) guarantee that the intrinsic and extrinsic parts of the orbital AM $\langle \mathbf{L} \rangle$ are transformed as $\langle \mathbf{L}^{\text{int}} \rangle \rightarrow \langle \mathbf{L}^{\text{int}} \rangle$ and $\langle \mathbf{L}^{\text{ext}} \rangle \rightarrow \langle \mathbf{L}^{\text{ext}} \rangle + \mathbf{r}_0 \times \langle \mathbf{P} \rangle$. One can say that the intrinsic orbital AM is the orbital AM calculated with respect to the centroid of the field, i.e., when $\langle \mathbf{r} \rangle = 0$. It should be remarked, however, that the definition (4.5) depends on the particular definition of the field centroid (mean coordinates) $\langle \mathbf{r} \rangle$. The coordinates weighted with either the energy density, or the energy-flux density, or the photon-number density may result in different values of $\langle \mathbf{r} \rangle$, see examples in [44,117,118]. Nonetheless, they all coincide in paraxial monochromatic beams (2.15). Note that the above equations can be applied to optical beams localized in two transverse dimensions ($\langle \mathbf{r} \rangle = \langle \mathbf{r}_\perp \rangle$, longitudinal coordinate does not contribute to the extrinsic AM), as well as to wave packets localized in three dimensions.

Naturally, the main properties of the extrinsic orbital AM are similar to (I) for mechanical particles and are in sharp contrast to the intrinsic orbital AM (IV):

> **Shifted beam AM:** Orbital, Extrinsic, Transverse (out-of-plane). Key parameters: $\langle \mathbf{r} \rangle, \langle \mathbf{k} \rangle$. (X)

Here we indicated the "out-of-plane" geometry, because the AM $\langle L_x^{\text{ext}} \rangle = \langle y \rangle \langle P_z \rangle$ (as shown in Fig. 17) is defined for the $y$-localized and $z$-propagating beams formed by wave vectors with different $(k_y, k_z)$ components [64].

4.1.2. *Extrinsic AM in the spin Hall effect of light.* It might seem at first that the extrinsic orbital AM does not make a physical difference, and cannot produce any observable effects. Indeed, the coordinate origin is an *abstract* object, and an extrinsic quantity cannot play any role in free space. However, any interaction with matter (which singles out some specific coordinate frame attached to it) can involve the extrinsic orbital AM in observable phenomena. The most remarkable example is the *spin-Hall and orbital-Hall effects* of light [63–73]. This group of phenomena stems from the spin-orbit interactions of light [31,173–175] and appears as transverse spin- and vortex-dependent shifts of light (see also [122,176–185]). These effects attracted rapidly growing attention during the past decade (see [174,175] for reviews), and here we only consider the simplest model example of the optical spin Hall effect.

We examine the total reflection of a paraxial optical beam (2.15) (for simplicity, without a vortex: $\ell = 0$) at a planar interface between free space and an isotropic non-absorbing medium, Fig. 18. The natural laboratory coordinates $(X, y, Z)$ are attached to the interface $Z = 0$, such that the medium occupies the $Z > 0$ half-space. The beam impinges the interface at an angle $\theta$ with respect to the $Z$-axis, and its axis lies in the $(X, Z)$ plane, as shown in Fig. 18. From the law of reflection, the propagation directions of the incident and reflected beams can be characterized by the following unit vectors:

$$\bar{\mathbf{z}} = \bar{\mathbf{Z}} \cos\theta + \bar{\mathbf{X}} \sin\theta \quad \text{and} \quad \bar{\mathbf{z}}' = -\bar{\mathbf{Z}} \cos\theta + \bar{\mathbf{X}} \sin\theta. \quad (4.6)$$

Hereafter we use primes to indicate quantities related to the reflected beam.

According to Eqs. (2.18), the incident and reflected beams carry longitudinal momenta and spin AM given by

$$\langle \mathbf{P} \rangle \propto k\bar{\mathbf{z}}, \quad \langle \mathbf{S} \rangle \propto \sigma\bar{\mathbf{z}} \quad \text{and} \quad \langle \mathbf{P}' \rangle \propto k\bar{\mathbf{z}}', \quad \langle \mathbf{S}' \rangle \propto \sigma'\bar{\mathbf{z}}'. \quad (4.7)$$

Here $\sigma$ and $\sigma'$ are the polarization helicities of the two beams, and we omit inessential factors, which do not affect the considerations below. The incident-beam helicity $\sigma$ is given by the initial conditions, whereas the helicity of the reflected beam depends on the properties of the particular interface: e.g., $\sigma' = -\sigma$ for a reflection from an ideal metal and $\sigma' \simeq \sigma$ for the total internal reflection at a dielectric interface near the critical incidence. Below we use only general arguments, which are independent of the particular medium.

Note that the medium under consideration and the corresponding Maxwell equations describing light are *rotationally symmetric* with respect to the $Z$-axis. Hence, according to the Noether theorem, the $Z$-component of the total angular momentum should be conserved in time. Furthermore, the medium plays the role of an external reflecting scalar potential in Maxwell equations, and it does not acquire any AM. Therefore, the $Z$-component of the total AM of light must be conserved: $\langle J_Z \rangle = \text{const}$. Next, we can substitute the continuous reflection of an infinitely-long stationary beam with the reflection of an arbitrarily long but finite wave packet. Then, only the incident (reflected) beam-packet exists at time $t = -\infty$ ($t = +\infty$), and we conclude that the *Z-component of the total AM of the incident beam-packet must be equal to the Z-component of the total AM of the reflected beam-packet*: $\langle J_Z \rangle = \langle J'_Z \rangle$. Obviously, the dynamical characteristics of arbitrary long paraxial wave packets are the same as for the beams (where the integral values (2.18) are calculated per unit propagation length).

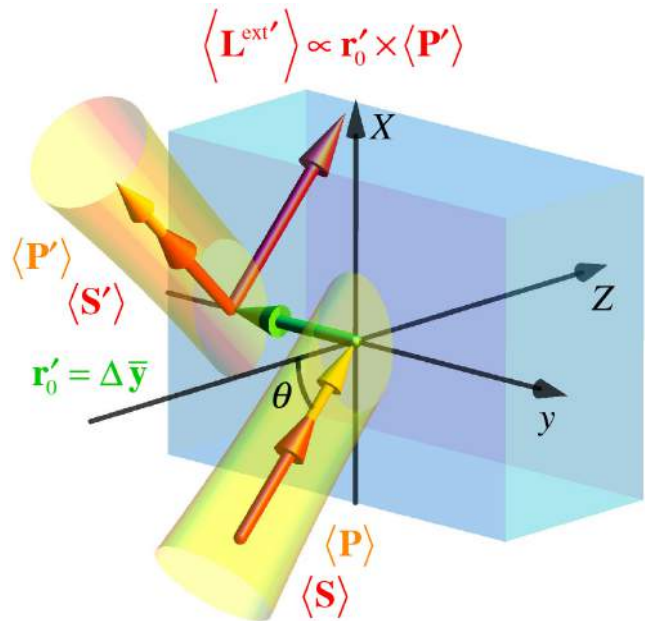

**Fig. 18.** Schematics of the spin Hall effect of light (transverse beam shift) at a planar interface. The reflection of a circularly-polarized light beam at a planar interface produces a transverse spin-dependent beam shift (4.10) [63,64,66,68,73]. In the simplest case of total reflection and isotropic non-absorbing media, this shift can be derived solely from the balance of the $z$-component of the total AM in the system [61–64,73], Eq. (4.9). The spin-Hall shift generates a transverse extrinsic orbital AM in the reflected beam (Fig. 17), which ensures the conservation of the $z$-component of the total AM between the incident and reflected light.

However, projecting the longitudinal spin AM of the incident and reflected beams, Eqs. (4.6) and (4.7) on the $Z$-axis, we have: $\langle S_Z \rangle \propto \sigma \cos\theta$ and $\langle S'_Z \rangle \propto -\sigma' \cos\theta \neq \langle S_Z \rangle$. This seemingly contradicts the AM conservation law in the problem. In fact, this difference between the $Z$-components of the spin AM of the incident and reflected light can be compensated from only one source: the *extrinsic orbital AM* (4.3) and (4.4). To produce this AM, the reflected beam must be *shifted along the y axis* by some distance $\mathbf{r}_0 = \Delta \bar{\mathbf{y}}$, as shown in Fig. 18. Then, this beam will possess the orbital AM

$$\langle \mathbf{L}' \rangle \equiv \langle \mathbf{L}^{\text{ext}\prime} \rangle \propto \mathbf{r}_0 \times \langle \mathbf{P}' \rangle. \tag{4.8}$$

Projection onto the $Z$-axis using Eqs. (4.6) and (4.7) yields $\langle L_Z^{\text{ext}\prime} \rangle \propto -\Delta k \sin\theta$. Now we can satisfy the above AM conservation law:

$$\langle S_Z \rangle = \langle S_Z' \rangle + \langle L_Z^{\text{ext}\prime} \rangle, \tag{4.9}$$

when

$$\Delta = -(\sigma + \sigma')\frac{\cot\theta}{k}. \tag{4.10}$$

Equation (4.10) describes the simplest case of the spin Hall effect of light, also known as the Imbert–Fedorov transverse shift [63,64,66,68] (see [73] for a review). It means that a beam of light carrying intrinsic longitudinal spin AM experiences a spin-dependent transverse shift (4.10) after reflection (or refraction) at a plane interface. This shift was predicted by Fedorov in 1955 and detected for the first time by Imbert in 1972 [186,187]. However, the original theoretical explanation of this effect was misleading, and the 50-years-long studies of this fine phenomenon were full of controversies. Only recently the effect was properly explained in terms of the spin-orbit interactions of light [63,64,68,73] and measured with a great accuracy using the "quantum weak measurement" technique [66,72,188–193] (see Fig. 19 for examples of experiments measuring the spin Hall effect of light). Although the shift (4.10) is small (a fraction of the wavelength), it is important both because of its fundamental nature and considerable contribution at the nano-scales of modern optics [72,122,175,182].

The role of the AM conservation (4.9) and the extrinsic orbital AM in the transverse beam shift was first revealed by Player and Fedoseyev in 1987 [61,62], and later confirmed in [63,64,69,70]. This could be regarded as the first example indicating the importance of the transverse extrinsic orbital AM of light for observable effects. Note that here we considered the simplest total-reflection case and derived the exact expression for the shift (4.10) from heuristic AM-balance considerations. At the same time, the proper solution of the generic beam-reflection/refraction problem requires rather delicate calculations involving the Fourier spectra of the beams [64,68,73] or the corresponding three-dimensional real-space fields [194]. But in any case the exact solutions of the beam reflection/refraction problems at planar isotropic interfaces satisfy the total AM conservation law [70,73].

We also briefly mention two important extensions of the transverse spin-dependent beam shift considered above. First, a quite similar *vortex*-dependent shift and *orbital* Hall effect takes place for paraxial beams carrying intrinsic longitudinal orbital angular momentum (2.18) $\langle \mathbf{L}^{\text{int}} \rangle \propto \ell \langle \mathbf{k} \rangle / k$. In this case, the balance between the intrinsic and extrinsic parts of the orbital AM results in observable beam-shift effects akin to Eqs. (4.9) and (4.10) [69–71,73,179,181]. Second, instead of *sharp interfaces* with jumps in the propagation direction of light, one can consider the *smooth* propagation of light in a *gradient-index* dielectric medium. The evolution of light in such medium is described by the Hamiltonian equations of motion for coordinates $\langle \mathbf{r} \rangle$ and momentum $\langle \mathbf{P} \rangle$, and a smooth trajectory, exactly as for a particle in classical mechanics [195]. Taking into account the spin-orbit interaction in Maxwell equations with "semi-classical" wavelength-order corrections [196], one can derive the helicity-dependent corrections to the traditional geometrical-optics equations for the trajectory of light [63,65,67,173,180]. The resulting equations have fundamental importance for the evolution of various spinning particles [197–199]. Importantly, these equations of motion possess a new integral of motion – the total angular momentum of light including the intrinsic spin and extrinsic orbital contributions [63,67,196]:

$$\left\langle \mathbf{L}^{\text{ext}} \right\rangle + \left\langle \mathbf{S} \right\rangle = \left\langle \mathbf{r} \right\rangle \times \left\langle \mathbf{P} \right\rangle + \sigma \frac{\left\langle \mathbf{P} \right\rangle}{k} = \text{const}. \qquad (4.11)$$

Figure 19 shows examples of experimental observations of the spin Hall effect of light and the corresponding beam shifts in various physical situations: beam refraction at a dielectric interface [66], propagation of light along a smooth curvilinear trajectory [67], and refraction into surface-plasmon beams at a light-plasmon interface [72].

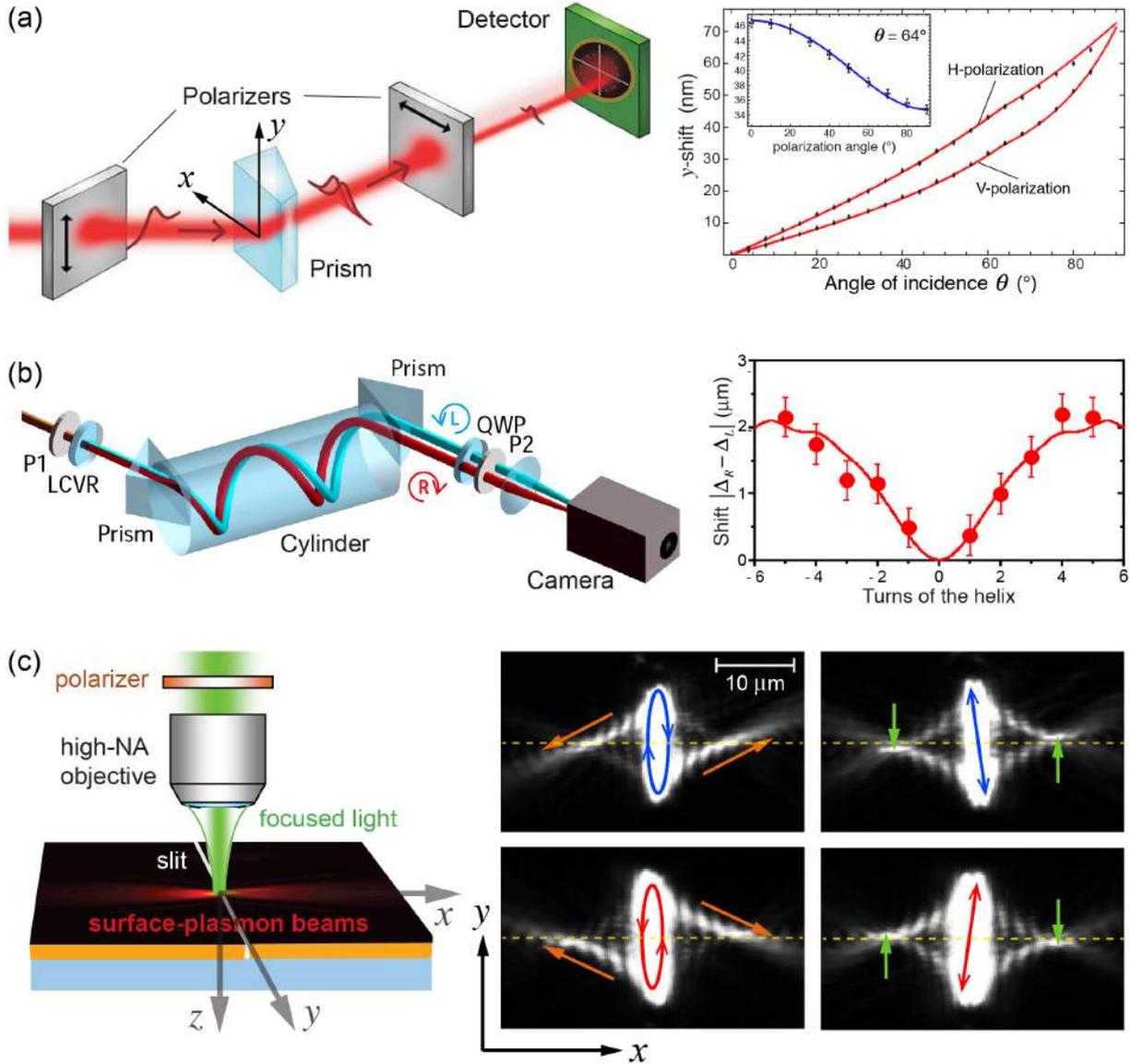

**Fig. 19.** Examples of experiments [66,67,72] demonstrating the transverse spin-dependent shifts (spin Hall effect) of light, which are related to the AM conservation as in Eqs. (4.9) and (4.10). **(a)** A tiny subwavelength shift in the beam refraction was amplified in [66] using the "quantum weak measurements" technique [188–193] and, as a result, measured with extraordinary Ångstrom accuracy. **(b)** The smooth propagation of light along a curvilinear (helical here) trajectory also brings about a transverse spin-dependent deflection [67], which ensures the conservation of the total AM of light (4.11). **(c)** Spin-Hall effect with surface-plasmon beams generated at a single slit on the metal surface [72]. An amplification by the "quantum weak measurement" method makes the beam shifts visible, in both momentum [directions shown by orange arrows] and coordinates [shown by green arrows].

An interesting extension of the spin Hall effect of light was suggested by Aiello *et al.* in [44], where they considered the usual paraxial beam (2.15) carrying a longitudinal spin AM (2.18): $\langle \mathbf{S} \rangle \propto \sigma \langle \mathbf{k} \rangle / k$. Instead of real changes in the direction of propagation of the beam (as in refraction or reflection), Aiello *et al.* considered "virtual" rotation of the coordinate frame by the angle $\theta$ in the $(x,z)$-plane in free space. In the new $(X, y, Z)$ coordinates, the longitudinal spin AM of the beam acquires a "transverse" component, $\langle S_X \rangle = -\langle S_z \rangle \sin\theta \neq 0$, while the centroid of the Poynting-momentum flux through a tilted $(X, y)$ plane becomes shifted: $\langle y \rangle_{\Pi_z} = \sigma \tan\theta / 2k$ [cf. Eq. (4.10)]. This effect was called the "geometric spin-Hall effect of light". Naturally, the rotation of a coordinate frame cannot change any beam properties in free space, and this shift is also "virtual". It is related to the definition of the beam centroid via a flux of the tilted Poynting vector. At the same time, the beam centroid defined via the energy density remains non-shifted: $\langle y \rangle_W = 0$ [44]. The angular momentum of the beam also remains purely *longitudinal*, as in the frame-independent Eq. (2.18). Nonetheless, the non-zero shift $\langle y \rangle_{\Pi_z}$ can play a role in the interaction of light with oblique detectors sensitive to the momentum density [200,201].

*4.2. Transverse intrinsic orbital AM*

In paraxial monochromatic vortex beams (2.15) the intrinsic orbital AM is always longitudinal, Eq. (2.18): $\langle \mathbf{L}^{int} \rangle \propto \ell \langle \mathbf{P} \rangle / k$. Moreover, one can show that the general definition (4.5) result in $\langle \mathbf{L}^{int} \rangle \| \overline{\mathbf{z}}$ when $\langle \mathbf{P} \rangle \| \overline{\mathbf{z}}$ and the beam centroid is determined as the transverse centroid of the longitudinal momentum density $P_z$: $\langle \mathbf{r} \rangle = \langle \mathbf{r}_\perp \rangle_{P_z}$. However, for the generic optical field and more natural energy-centroid definition $\langle \mathbf{r} \rangle = \langle \mathbf{r} \rangle_W$ the intrinsic orbital AM (4.5) is not restricted to be purely longitudinal. Can there be optical fields *with non-collinear intrinsic orbital AM and momentum*, $\langle \mathbf{L}^{int} \rangle \nparallel \langle \mathbf{P} \rangle$, i.e., a non-zero transverse component $\langle \mathbf{L}^{int}_\perp \rangle$? This question was recently examined by Bliokh and Nori [47], and it was concluded that such transverse intrinsic orbital AM can naturally occur in *polychromatic* fields with non-stationary intensity distributions.

Since the intrinsic orbital AM of paraxial beams is associated with optical vortices inside the beam, one can expect the transverse intrinsic orbital AM in field with *transverse vortices*. In contrast to the *screw* dislocation of phase fronts in longitudinal vortices (Fig. 2b), such transverse or skew vortices represent *edge* or mixed *edge-screw* dislocations in phase fronts [39,41,43]. In other words, the vortex singularity line, which is the nodal intensity line, is skew or orthogonal to the propagation direction of the wave. Note that transverse vortex-like circulations of the Poynting vector occur in nonparaxial light fields in many basic interference and diffraction problems [202–206]. However, in all these problems, transverse vortices appear either in vortex-antivortex pairs or in non-localized fields where the integral AM is ill-defined. Furthermore, previous studies [202–206] found such transverse circulations in the *Poynting* vector, and no proper AM analysis based on the *canonical* momentum (2.9) has been made. Thus, so far no transverse intrinsic orbital AM has been found in monochromatic optical fields. At the same time, skew vortices and intrinsic orbital AM are ubiquitous in polychromatic fields with transversely-moving intensity distributions [47].

Since in this Section 4.2 we discuss only orbital AM, for simplicity we will ignore polarization (spin) degrees of freedom and consider *scalar* waves. To construct non-collinear momentum and intrinsic orbital AM, $\langle \mathbf{L}^{int} \rangle \nparallel \langle \mathbf{P} \rangle$, we look for *non-stationary* wave fields with *transversely moving vortices*. Such solutions are very easy to find in free space. It is sufficient to observe the usual stationary vortex beam (2.15) in a *transversely-moving reference frame*. Indeed, let the observer moves with velocity $\mathbf{v} = v\overline{\mathbf{x}}$ along the *x* axis. The vortex line (and, hence, the intrinsic orbital AM) will keep its direction along the *z*-axis: $\langle \mathbf{L}^{int} \rangle = \langle L^{int} \rangle \overline{\mathbf{z}}$, i.e., orthogonal to the observer motion. At the same time, the beam will become non-stationary and moving with the velocity $-\mathbf{v}$ along the $x$-axis. Thus, the beam will acquire momentum component along the $x$-axis, $\langle P_x \rangle$, which is transverse with respect to the intrinsic orbital AM.

To quantify this consideration, we now consider the Lorentz transformation of the momentum and angular momentum of a particle in the relativistic mechanics. In the original reference frame $(t, \mathbf{r})$, let the particle have energy $E$, momentum $\mathbf{p} = p\bar{\mathbf{z}}$, and AM $\mathbf{L} = L\bar{\mathbf{z}}$. Then, the Lorentz boost with $\mathbf{v} = v\bar{\mathbf{x}}$ results in the following energy, momentum, and AM values in the moving reference frame $(t', \mathbf{r}')$ [112,207,208]:

$$E' = \gamma E, \quad \mathbf{p}' = \mathbf{p} - \gamma \frac{E}{c^2}\mathbf{v}, \quad \mathbf{L}' = \gamma \mathbf{L}, \tag{4.12}$$

where $\gamma = 1/\sqrt{1 - v^2/c^2}$ is the Lorentz factor. Thus, the momentum and AM become non-parallel to each other due to the transverse momentum acquired by the particle. The difference in transformations of the momentum and AM in relativity is explained by the fact that the momentum is part of the energy-momentum *four-vector*, while the AM is part of the *antisymmetric rank-2 AM tensor* [207,208].

Considering now the same Lorentz boost for a scalar paraxial wave beam similar to Eq. (2.15) (see Figs. 20 and 21):

$$\psi(\mathbf{r},t) \simeq A(\rho,z)\exp(ikz + i\ell\varphi - i\omega t). \tag{4.13}$$

The beam is stationary in the original reference frame $(t, \mathbf{r})$ and is characterized by the energy, momentum, and intrinsic orbital AM similar to Eq. (2.18):

$$\langle W \rangle \propto \omega, \quad \langle \mathbf{P} \rangle = \frac{\langle W \rangle}{\omega} k\bar{\mathbf{z}}, \quad \langle \mathbf{L} \rangle = \frac{\langle W \rangle}{\omega}\ell\bar{\mathbf{z}}. \tag{4.14}$$

Performing the Lorentz transformation of the field (4.13) and calculating the same quantities in the moving reference frame $(t', \mathbf{r}')$, one can obtain [112]

$$\langle W' \rangle \propto \gamma\omega, \quad \langle \mathbf{P}' \rangle = \frac{\langle W \rangle}{\omega}\left(k\bar{\mathbf{z}}' - \gamma\frac{\omega}{c^2}\mathbf{v}\right), \quad \langle \mathbf{L}' \rangle = \frac{\langle W \rangle}{\omega}\gamma\ell\bar{\mathbf{z}}', \tag{4.15}$$

which is in agreement with Eqs. (4.12).

There is a subtle issue, however, and the orbital AM $\langle \mathbf{L}' \rangle$ is not purely intrinsic anymore. Careful consideration shows that the beam acquires a *transverse AM-dependent shift* [112,209]:

$$\mathbf{r}'_0 = -\frac{\mathbf{v} \times \langle \mathbf{L} \rangle}{2\langle W \rangle} - \mathbf{v}t', \tag{4.16}$$

which is defined here as a centroid of the photon-number density (the transverse displacement of the energy-density centroid being twice as large). This observer- and AM-dependent transverse shift can be called "relativistic Hall effect" [112,113,210–212] (Fig. 21). Notably, the analogous spin-AM-dependent shift for relativistic particles is related to the conservation of the total AM, similar to the spin-Hall effect in Section 4.1.2 [210]. Thus, the orbital AM (4.15) consists of the following intrinsic and extrinsic contributions:

$$\langle \mathbf{L}^{\text{int}'} \rangle = \frac{\langle W \rangle}{\omega}\frac{\gamma + \gamma^{-1}}{2}\ell\bar{\mathbf{z}}', \quad \langle \mathbf{L}^{\text{ext}'} \rangle = \mathbf{r}'_0 \times \langle \mathbf{P}' \rangle, \tag{4.17}$$

which ensure that $\langle L_z^{\text{int}'} \rangle + \langle L_z^{\text{ext}'} \rangle = \langle L_z' \rangle$, in agreement with Eq. (4.15).

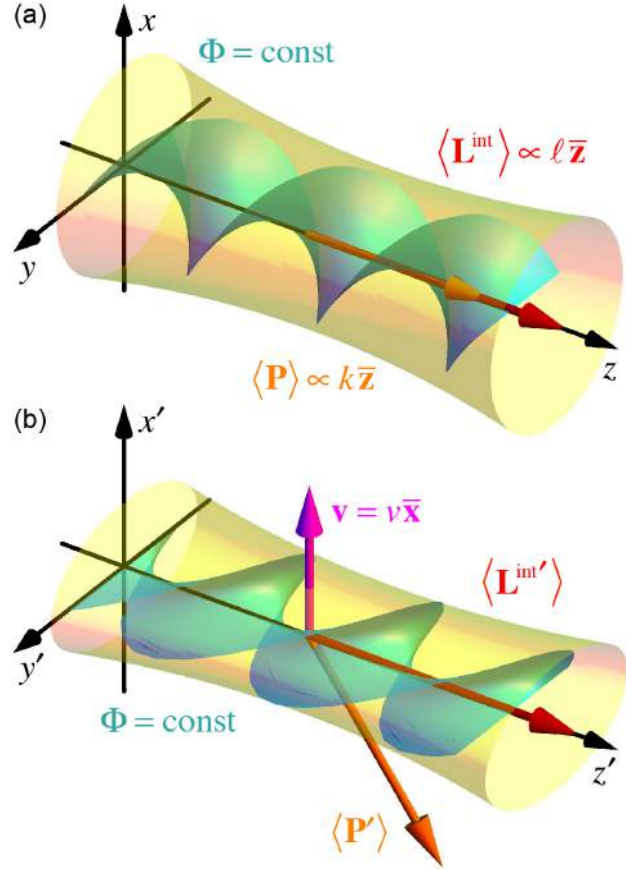

**Fig. 20.** Transverse Lorentz boost of the paraxial vortex beam carrying an intrinsic orbital AM [47,112]. **(a)** Monochromatic vortex beam (as in Fig. 2b), in the motionless reference frame $(t,\mathbf{r})$, with the longitudinal intrinsic orbital AM (2.18) $\langle \mathbf{L}^{int}\rangle$. **(b)** The same beam in the reference frame $(t',\mathbf{r}')$ moving with the transverse velocity $\mathbf{v} = v\bar{\mathbf{x}}$. The beam becomes non-stationary (polychromatic) and acquires the transverse momentum component $\langle P'_x\rangle = -\gamma k \frac{v}{c}$, Eq. (4.15). Thus, the beam momentum $\langle \mathbf{P}'\rangle$ and intrinsic orbital AM $\langle \mathbf{L}^{int\prime}\rangle$, Eq. (4.17), become non-collinear to each other, and the beam carries a *transverse intrinsic orbital AM component* (4.19) and (XI). The tilted phase fronts of the vortex signify the mixed edge-screw phase dislocation in the transversely-moving vortex, i.e., *spatio-temporal* vortex [47,213] (see Fig. 21).

Equations (4.15) and (4.17) reveal the *non-collinear* momentum $\langle \mathbf{P}'\rangle$ and intrinsic orbital AM $\langle \mathbf{L}^{int\prime}\rangle$ of the beam, Figs. 20 and 21. Let us introduce the coordinate frame $(X, y, Z)$ rotated by the angle $\theta = \arctan\left(\gamma \frac{v}{c}\right)$ in the $(x,z)$-plane, such that the beam momentum is aligned with the $Z$-axis [113]:

$$\bar{\mathbf{z}}' = \bar{\mathbf{Z}}\cos\theta + \bar{\mathbf{X}}\sin\theta \quad \text{and} \quad \bar{\mathbf{x}}' = -\bar{\mathbf{Z}}\sin\theta + \bar{\mathbf{X}}\cos\theta. \tag{4.18}$$

Using $\sin\theta = \frac{v}{c}$, $\cos\theta = \gamma^{-1}$, and $\omega = kc$, we obtain the momentum (4.15) and intrinsic orbital AM (4.17) in these coordinates:

$$\langle \mathbf{P}'\rangle \propto \gamma k \bar{\mathbf{Z}}, \quad \langle \mathbf{L}^{int\prime}\rangle \propto \frac{\gamma + \gamma^{-1}}{2}\gamma^{-1}\ell\bar{\mathbf{Z}} + \boxed{\frac{\gamma + \gamma^{-1}}{2}\frac{v}{c}\ell\bar{\mathbf{X}}}. \tag{4.19}$$

Here, the last term in the orange frame is the *transverse intrinsic orbital AM* $\langle \mathbf{L}_\perp^{\text{int}'} \rangle$, induced by the transverse Lorentz boost of the beam. Note that this AM lies in the plane formed by $\langle \mathbf{P} \rangle$ and $\mathbf{v}$, and it appears in linear order in $v/c$. Summarizing the properties of the transverse AM in Eq. (4.19):

> **Lorentz-boosted beam AM:** Orbital, Intrinsic, Transverse (in-plane).
> Key parameters: $\ell, \mathbf{v}$. (XI)

Note that here the boost velocity $\mathbf{v}$ substituted the beam momentum $\langle \mathbf{k} \rangle$ in the longitudinal intrinsic orbital AM (IV). Since the velocity and momentum share the same $\mathcal{P}$-odd and $\mathcal{T}$-odd nature, Eq. (4.19) have the proper $\mathcal{T}$-odd symmetry of the AM.

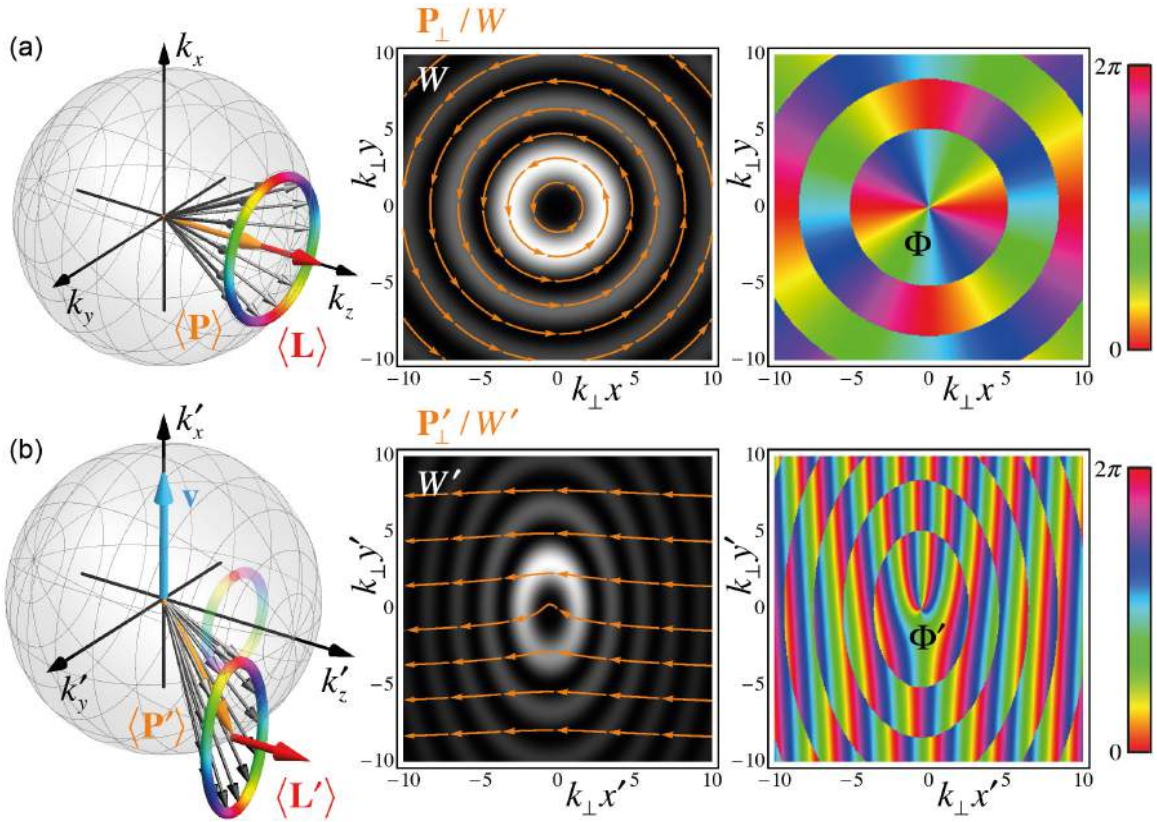

**Fig. 21.** Transverse Lorentz boost of the scalar Bessel beam (4.13) with $A(\rho, z) = J_{|\ell|}(k_\perp \rho)$ and $k \to k_z$, where $k_\perp = \sqrt{k^2 - k_z^2}$. Here $k_\perp / k = 0.5$ and $\ell = 2$, while the Lorentz transformation is characterized by $v/c = 0.8$. **(a)** The original beam's Fourier spectrum (in $\mathbf{k}$-space), its transverse intensity and momentum-density distributions (as in Fig. 3b), as well as the transverse phase distribution showing a screw phase dislocation (vortex). **(b)** The same beam observed in a transversely-moving reference frame with velocity $\mathbf{v} = v\bar{\mathbf{x}}$. The beam spectrum is now shifted away from the iso-frequency sphere, i.e., the beam becomes *polychromatic*. The transverse motion of the vortex (clearly seen in the transverse-momentum distribution) makes its momentum $\langle \mathbf{P}' \rangle$ and orbital AM $\langle \mathbf{L}' \rangle$ tilted with respect to each other, Eq. (4.19). The transverse phase distribution exhibits a mixed edge-screw dislocation, which signifies the *spatio-temporal* character of the vortex [47,213]. The transverse intensity distribution reveals the $x$-squeezing of the beam due to the Lorentz contraction and also a small transverse $y$-deformation due to the transverse shift of beam's centroid (4.16), i.e., the relativistic Hall effect [112,113].

It is worth remarking that the transversely-moving beam carries a *spatio-temporal* vortex [47,213]. It is characterized by a mixed edge-screw dislocation in phase [39,41,43] (see Figs. 20b and 21b), and forms a vortex and circulating four-current not only in the $(x,y)$ or $(k_x,k_y)$ planes but also in the $(t,y)$ or $(\omega,k_y)$ planes [47]. Figure 21 shows details of the transformation of the scalar vortex Bessel beam (4.13) observed in a transversely Lorentz-boosted reference frame. Transversely-moving spatio-temporal vortices carrying transverse intrinsic orbital AM component (XI) can appear in two-dimensional surface waves (e.g., surface plasmon-polaritons), nonlinear media [213], or in interference of three or more plane waves with different frequencies (cf. [214]). In addition, such vortex states of light can naturally be generated from moving sources: e.g., astrophysical ones [215–217].

## 5. Conclusions

In this review we aimed to provide a unified theory of the angular momentum of light (Section 2) and a comprehensive picture of various types of spin and orbital AM in structured optical fields (Sections 3 and 4). We also provided an overview of the main experiments revealing the traditional and novel types of optical AM.

In contrast to previous reviews on the optical AM [5–13], we considered the *canonical* picture of the optical AM, which is based on the canonical momentum and spin densities rather than the Poynting vector. Such description is in perfect agreement with experiments, free of difficulties appearing with the Poynting momentum and AM, and it naturally admits separation of orbital and spin degrees of freedom of light.

Most importantly, the generic theory enabled us to thoroughly investigate and classify novel types of spin and orbital AM, which are attracting rapidly-growing attention. In addition to the most common longitudinal spin and orbital AM in paraxial beams (previously explored in detail in many books and reviews [5–13]), we described a number of *transverse* AM of different nature. The rather unusual and contrasting properties of different AM kinds are summarized in Eqs. (I)–(XI) throughout the review and compiled here in Table I. One can see that various kinds of optical AM reveal dependencies on a variety of key parameters. This is in contrast to the traditional picture with the longitudinal spin AM determined by the wave helicity and the longitudinal orbital AM determined by the vortex charge.

Apparently, the most intriguing new kind of AM are the *transverse spin AM* (Section 3). We have revealed at least three basic types of transverse spin:

- The transverse (out-of-plane) helicity-independent spin AM density in evanescent and interfering propagating waves. This spin AM is determined solely by the wave-vector and phase parameters of the waves, and it is orthogonal to the wave vectors in the field.

- The transverse (in-plane) "dual-antisymmetric" spin AM density in evanescent and interfering and propagating waves. This spin AM has opposite "electric" and "magnetic" parts, so that the total spin vanishes, but these parts can appear in dual-asymmetric light-matter interactions. This spin lies in the plane with the wave-vectors of the field but orthogonally to the main momentum. Furthermore, it is proportional to the second Stokes parameter of the polarization rather than the helicity.

- The integral transverse (in-plane) spin AM, which appears in the interference of propagating waves with different wave vectors and helicities. This integral spin is generated by the vector sum of the usual longitudinal spins of the interfering waves, and, hence, depends on their helicities.

The first of the above, the transverse helicity-independent spin AM, was recently measured in several experiments in both propagating and evanescent fields. Most importantly, it promises remarkable applications in spin-controlled unidirectional interfaces with evanescent waves. This is because such transverse spin in evanescent waves is intimately related to fundamental topological properties of Maxwell equations in free space.

We have also described the *transverse orbital AM* of light (Section 4). Here one should distinguish the intrinsic and extrinsic types of the integral orbital AM:

- The extrinsic orbital AM is the most common and trivial type of AM, which is entirely analogous to that of a classical point particle. Nonetheless, it can play a crucial role in observable optical effects involving spin-orbit interactions and the conservation of the total AM. We have shown that the transverse extrinsic orbital AM is responsible for the spin-Hall effect of light refracted or reflected in inhomogeneous media.

- The transverse part of the intrinsic orbital AM (from the optical vortex) can appear only in polychromatic fields. In particular, it naturally appears in Lorentz-boosted beams observed in a transversely-moving reference frame.

Notably, the transverse extrinsic orbital AM and the transverse helicity-independent spin in evanescent waves are closely related to different manifestations of the *spin-orbit interactions* of light. Namely, they underpin the spin Hall effect (i.e., small spin-dependent transverse shift) and quantum spin Hall effect of light (i.e., spin-dependent unidirectional edge modes), respectively. The spin-orbit interactions of light originate from the transversality of electromagnetic waves and the corresponding longitudinal field components in the non-plane-wave fields. As we have seen in Section 3, these longitudinal field components are also responsible for the generation of the transverse spin AM.

To summarize, we have shown how a very concise and fundamental theory of the spin and orbital AM of light in generic optical fields describes a rich variety of AM forms with extraordinary features in specific light configurations. All of these angular momenta appear from the interplay between the *particle* (localized) and *wave* (extended) features of *structured* light fields. Modern nano-optics and photonics tend to explore and employ new degrees of freedom of structured light, which are absent in simple plane waves or particle-like Gaussian wave packets. Therefore, these frontier areas of research offer extraordinary opportunities for exploiting longitudinal and transverse angular momenta of light in various fundamental studies and applications.

## Acknowledgements

We acknowledge fruitful discussions and correspondence with A.Y. Bekshaev, A. Aiello, P. Banzer, P. Schneeweiss, E.A. Ostrovskaya, R. Mathevet, G.L.J.A. Rikken, and T. Kawalec. This work was partially supported by the RIKEN iTHES Project, MURI Center for Dynamic Magneto-Optics (grant no. FA9550-14-1-0040), JSPS-RFBR (contract no. 12-02-92100), a Grant-in-Aid for Scientific Research (A), and the Australian Research Council.

| | System and AM type | Density | Integral | Ext./ Int. | Key parameters |
|---|---|---|---|---|---|
| **Particle/ wave** | **Point particle**: Transverse orbital AM | | $\mathbf{L} = \mathbf{r} \times \mathbf{p}$ | ext. | $\mathbf{r}$, $\mathbf{p}$ |
| | **Plane wave**: Longitudinal spin AM | $\mathbf{S} \propto \sigma \dfrac{\mathbf{k}}{k}$ | | int. | $\sigma$, $\mathbf{k}$ |
| **Paraxial monochrom. beams** | **Polarized beam**: Longitudinal spin AM | $\mathbf{S} \simeq \sigma \dfrac{\mathbf{P}_\parallel}{k}$ | $\langle \mathbf{S} \rangle \simeq \sigma \dfrac{\langle \mathbf{P} \rangle}{k}$ | int. | $\sigma$, $\langle \mathbf{k} \rangle$ |
| | **Vortex beam**: Transv./Long. orbital AM | $\mathbf{L} = \mathbf{r} \times \mathbf{P}$ | $\langle \mathbf{L} \rangle \simeq \ell \dfrac{\langle \mathbf{P} \rangle}{k}$ | ext./ int. | $\ell$, $\langle \mathbf{k} \rangle$ |
| **Other structured fields** | **Evanesc. wave**: Transverse (out-of-plane) $\sigma$-indep. spin AM | $\mathbf{S}_\perp \propto \dfrac{\mathrm{Re}\,\mathbf{k} \times \mathrm{Im}\,\mathbf{k}}{(\mathrm{Re}\,\mathbf{k})^2}$ | $\langle \mathbf{S}_\perp \rangle^+ \propto \dfrac{\mathrm{Re}\,\mathbf{k} \times \mathrm{Im}\,\mathbf{k}}{(\mathrm{Re}\,\mathbf{k})^2}$ | int. | $\mathrm{Re}\,\mathbf{k}$, $\mathrm{Im}\,\mathbf{k}$ |
| | **Evanesc. wave**: Transverse (in-plane) anti-dual spin AM | $\mathbf{S}_\perp^{e,m} \propto \pm\chi \dfrac{k\,\mathrm{Im}\,\mathbf{k}}{(\mathrm{Re}\,\mathbf{k})^2}$ | $\langle \mathbf{S}_\perp^{e,m} \rangle^+ \propto \pm\chi \dfrac{k\,\mathrm{Im}\,\mathbf{k}}{(\mathrm{Re}\,\mathbf{k})^2}$ | int. | $\chi$, $\mathrm{Im}\,\mathbf{k}$ |
| | **Interfering plane waves**: Transverse (out-of-plane) $\sigma$-indep. spin AM | $\mathbf{S}_\perp \propto \dfrac{\langle \mathbf{k} \rangle \times \delta\mathbf{k}}{k^2} \sin\delta\Phi$ | $\langle \mathbf{S}_\perp \rangle = 0$ | int. | $\langle \mathbf{k} \rangle$, $\delta\mathbf{k}$, $\delta\Phi$ |
| | **Interfering plane waves**: Transverse (in-plane) anti-dual spin AM | $\mathbf{S}_\perp^{e,m} \propto \mp\chi \dfrac{\delta\mathbf{k}}{k} \sin\delta\Phi$ | $\langle \mathbf{S}_\perp^{e,m} \rangle = 0$ | int. | $\chi$, $\delta\mathbf{k}$, $\delta\Phi$ |
| | **Interfering plane waves**: Transverse (in-plane) $\sigma$-dep. spin AM | … | $\langle \mathbf{S}_\perp \rangle = \delta\sigma \dfrac{\delta\mathbf{k}}{k}$ | int. | $\delta\sigma$, $\delta\mathbf{k}$ |
| **Modified paraxial beams** | **Shifted beam**: Transverse extrinsic orbital AM | | $\langle \mathbf{L}^{\mathrm{ext}} \rangle = \langle \mathbf{r} \rangle \times \langle \mathbf{P} \rangle$ | ext. | $\langle \mathbf{r} \rangle$, $\langle \mathbf{k} \rangle$ |
| | **Lorentz-boosted beam**: Skew intrinsic orbital AM | | $\langle \mathbf{L}_\perp^{\mathrm{int}} \rangle \propto \ell \dfrac{\mathbf{v}_\perp}{c}$, | int. | $\ell$, $\mathbf{v}$ |

**Table I.** Different types of the AM (I)–(XI), their basic features, and key parameters. Previous reviews [5–13] focused mostly on the paraxial monochromatic beams.